\documentclass[useAMS]{mn2e}
\usepackage{epsfig}

\def\herschel{{\sl Herschel}}
\def\inst#1{$^{#1}$}

\newcounter{subfigure}

\title[\herschel-ATLAS/GAMA clustering at low redshifts]{\herschel-ATLAS/GAMA: spatial clustering of low-redshift
sub-mm galaxies\thanks{\herschel\ is an ESA space observatory with science instruments provided
by European-led Principal Investigator consortia and with important participation from NASA}}
\author[van Kampen et al.]{E. van Kampen\inst{1}\thanks{e-mail: evkampen@eso.org},
 D.J.B. Smith\inst{2,13}, S. Maddox\inst{2,29}, A.M. Hopkins\inst{3}, I. Valtchanov\inst{4}, \newauthor 
 J.A. Peacock\inst{5}, M.J. Micha{\l}owski\inst{5}, P. Norberg\inst{6},
 S. Eales\inst{7},  L. Dunne\inst{2,29}, J. Liske\inst{1}, \newauthor
 M. Baes\inst{8}, D. Scott\inst{9},   E. Rigby\inst{2,5}, A. Robotham\inst{10},  P. van der Werf\inst{5,11},
 E. Ibar\inst{12},  \newauthor
 M.J. Jarvis\inst{13}, J. Loveday\inst{14}, 
 R. Auld\inst{5}, I.K. Baldry\inst{15}, S. Bamford\inst{2}, E. Cameron\inst{16},  \newauthor
 S. Croom\inst{17}, S. Buttiglione\inst{18}, A. Cava\inst{19},  A. Cooray\inst{20},  S. Driver\inst{10,21},
 J.S. Dunlop\inst{5}, \newauthor
 A. Dariush\inst{22},  J. Fritz\inst{8}, R.J. Ivison\inst{12}, 
 E. Pascale\inst{7}, M. Pohlen\inst{7}, G. Rodighiero\inst{25}, \newauthor
 P. Temi\inst{23},  D.G. Bonfield\inst{13},  D. Hill\inst{13},  D.H. Jones\inst{3}, L. Kelvin\inst{10},
 H. Parkinson\inst{5}, \newauthor 
 M. Prescott\inst{15},  R. Sharp\inst{3}, G. de Zotti\inst{18,24}, S. Serjeant\inst{26}, 
 C.C. Popescu\inst{27}, R.J. Tuffs\inst{28} \newauthor
\\
$^1$ European Southern Observatory, Karl-Schwarzschild-Strasse 2, D-85748 Garching bei M\"unchen, Germany\\
$^2$ School of Physics and Astronomy, University of Nottingham, University Park, Nottingham, NG7 2RD, UK\\
$^3$ Australian Astronomical Observatory, PO Box 296, Epping, NSW 1710, Australia\\
$^4$ Herschel Science Centre, ESAC, ESA, PO Box 78, Villanueva de la Ca\~nada, 28691 Madrid, Spain\\
$^5$ SUPA, Institute for Astronomy, University of Edinburgh, Royal Observatory,
Blackford Hill, Edinburgh EH9, 3HJ, UK \\
$^6$ Institute for Computational Cosmology, Department of Physics, Durham University, South Road, Durham DH1 3LE, UK\\
$^7$ School of Physics and Astronomy, Cardiff University, The Parade, Cardiff, CF24 3AA, UK\\
$^8$ Sterrenkundig Observatorium, Universiteit Gent, Krijgslaan 281 S9, B-9000 Gent, Belgium\\
$^9$ Department of Physics and Astronomy, 6224 Agricultural Road, University of British Columbia, Vancouver, BC, V6T 1Z1, Canada\\
$^{10}$ SUPA, School of Physics and Astronomy, University of St. Andrews, North Haugh, St. Andrews, KY16 9SS, UK\\
$^{11}$ Leiden Observatory, Leiden Observatory, P.O. Box 9513, NL-2300 RA Leiden, the Netherlands\\
$^{12}$ UK Astronomy Technology Center, Royal Observatory Edinburgh, Edinburgh, EH9 3HJ, UK\\
$^{13}$ Centre for Astrophysics Research, Science and Technology Research Institute, University of Hertfordshire, Herts AL10 9AB, UK\\
$^{14}$ Astronomy Centre, Department of Physics and Astronomy,
University of Sussex, Falmer, Brighton, BN1 9QH, UK\\
$^{15}$ Astrophysics Research Institute, Liverpool John Moores University,
      12 Quays House, Egerton Wharf, Birkenhead, CH41 1LD, UK\\
$^{16}$ ETH Zurich, Insitute for Astronomy, HIT J12.3, CH-8093 Zurich, Switzerland\\
$^{17}$ Sydney Institute for Astronomy, School of Physics, University of Sydney, NSW 2006, Australia\\
$^{18}$ INAF-Osservatorio Astronomico di Padova, Vicolo Osservatorio 5, I-35122 Padova, Italy\\
$^{19}$ Departamento de Astrof\'{\i}sica, Facultad de CC. F\'{\i}sicas, Universidad Complutense de Madrid, E-28040 Madrid, Spain\\
$^{20}$ Dept. of Physics \& Astronomy, University of California, Irvine, CA 92697, USA\\
$^{21}$ International Centre for Radio Astronomy Research (ICRAR), University of Western Australia, WA 6009\\
$^{22}$ Physics Department, Imperial College London, South Kensington campus, London, SW7 2AZ, UK\\
$^{23}$ Astrophysics Branch, NASA Ames Research Center, Mail Stop 245-6, Moffett Field, CA 94035, USA\\
$^{24}$ SISSA, Via Bonomea 265, I-34136 Trieste, Italy\\
$^{25}$ University of Padova, Vicolo Osservatorio 3, I-35122 Padova, Italy\\
$^{26}$ Dept. of Physics and Astronomy, The Open University, Milton Keynes, MK7 6AA, UK\\
$^{27}$ Jeremiah Horrocks Institute, University of Central Lancashire, Preston PR1 2HE, UK\\
$^{28}$ Max Planck Institute for Nuclear Astrophysics (MPIK), Saupfercheckweg 1, 69117 Heidelberg, Germany\\
$^{29}$ Department of Physics and Astronomy, University of Canterbury, Private Bag 4800, Christchurch, New Zealand\\
}

\begin{document}

\date{Accepted .... Received ....}

\pagerange{\pageref{firstpage}--\pageref{lastpage}} \pubyear{2012}

\maketitle

\label{firstpage}
\clearpage

\begin{abstract} 
We have measured the clustering properties of low-redshift ($z<0.3$) sub-mm galaxies detected
at 250 $\mu$m in the {\it Herschel}-ATLAS Science Demonstration Phase (SDP) field.
We selected a sample for which we have high-quality spectroscopic redshifts,
obtained from reliably matching the 250-$\mu$m sources to a complete (for $r<19.4$) sample
of galaxies from the GAMA database.
Both the angular and spatial clustering strength are measured for all $z<0.3$ sources as well
as for five redshift slices with thickness $\Delta z$=0.05 in the range $0.05<z<0.3$.
Our measured spatial clustering length $r_0$ is comparable
to that of optically-selected, moderately star-forming (blue) galaxies: we find values around 5 Mpc. 
One of the redshift bins contains an interesting structure, at $z=0.164$.
\end{abstract}


\section{Introduction}

A key statistical property of a given population of galaxies is its clustering length,
most often expressed as the scale-length of the galaxy-galaxy autocorrelation function. This statistic
quantifies the environment of such a galaxy population, and thus helps to interpret their other properties.
This paper deals with clustering of low-redshift sub-mm detected galaxies, which are expected to be
relatively normal galaxies, albeit with somewhat enhanced star formation as compared to high-redshift
sub-mm galaxies, which typically have much higher star formation rates and are thought to be
mostly merger-induced star-bursting galaxies. 

`Normal', local galaxies, as detected in the optical wavebands, have clustering lengths of
5--6 Mpc, depending on colour and/or luminosity (e.g. Coil et al. 2008 for the DEEP2 sample,
Zehavi et al. 2011 for the SDSS DR7 sample, and Christodoulou et al. 2012 for an analysis of the
SDSS DR7 sample using photometric redshifts calibrated using GAMA data).
In this paper we study the clustering properties of galaxies selected at 250 $\mu$m for which we also have
a spectroscopic redshift (of sufficient accuracy).

For sub-mm galaxies, various clustering estimates exist, most often for samples where redshift information
is sparse and only small fields are covered. Early attempts to measure the angular correlation function for
high-$z$ sub-mm galaxies detected with SCUBA include Almaini et al. (2003), Webb et al. (2003), Blain et al. (2004),
and Scott et al. (2006). A more recent estimate is that of Wei\ss\ et al. (2010), using LABOCA. In this paper we
make use of the Science Demonstration Phase (SDP) data of {\it Herschel}-ATLAS (Eales et al. 2010),
which offers a wider area than was available to the studies mentioned above, although at a shorter wavelength (and thus a
lower peak redshift). The angular clustering of all reliably detected 250 $\mu$m sources in this SDP field
has been measured by Maddox et al. (2010). A similar measurement for a similar sample, obtained as part of the
HerMES project, was presented by Cooray et al. (2010). Both these measurements have been compared to model 
predictions by Short \& Coles (2011).

Here we study a subset of the galaxies used for the clustering analysis of Maddox et al. (2010): those
with reliable spectroscopic redshifts. This allows us to study spatial clustering, although
only for redshifts below about 0.3, where we have sufficient numbers of redshifts.
In a related, complementary paper, Guo et al.\ (2011) measure the cross-correlation of a similar sample of galaxies,
to study the clustering bias of this sample and the properties of their haloes.

In Section 2 we first describe our particular sample of low-redshift sub-mm sources, and how it
was selected. In Section 3 we describe the methods to estimate angular and spatial clustering measures,
which are then applied in Section 4 to study the clustering properties of our sample.
In Section 5 we summarize our conclusions, and look forward to what we
can do once the full \herschel-ATLAS data set becomes available. In this paper we adopt the following cosmological
parameters where needed: $H_0$= 73 km s$^{-1}$ Mpc$^{-1}$, $\Omega_{\rm m}$=0.25, and $\Omega_{\Lambda}$=0.75.

\section{Observational data}

The observational data used come from a match of source catalogues obtained from \herschel-ATLAS (Eales et al. 2010)
and GAMA (Driver et al. 2011), performed by Smith et al.\ (2011). In this section we described the source
catalogues and selection. 

\subsection{\herschel-ATLAS data and SPIRE source catalogue}

Our sample of low-redshift sub-mm sources is extracted from the first \herschel-ATLAS data field that
was taken as part of the Science Demonstration Phase, as described in Eales et al.\ (2010).
\herschel-ATLAS is based on parallel scan mode observations performed with
{\it Herschel} (Pilbratt et al. 2010). Maps from the SPIRE (Griffin et al. 2010) data
were produced using a naive mapping technique after removing instrumental
temperature variations from the time-line data (Pascale et al. 2011).
Noise maps were generated by using the two
cross-scan measurements to estimate the noise per detector pass, and
then for each pixel the noise is scaled by the square-root of the number
of detector passes. 

Sub-mm sources were identified in the SPIRE maps as described in Rigby et al. (2011).
To produce a catalogue of reliable sources, only those that
are detected at the 5-$\sigma$ level at 250 $\mu$m were selected.
In calculating the $\sigma$ value for each source, the relevant noise
map was used, and the confusion noise was added to this in quadrature. The average
1-$\sigma$ instrumental noise values are 4, 4 and 5.7 mJy beam$^{-1}$
respectively in the 250, 350 and 500 $\mu$m bands. The
confusion noise was estimated from the difference between the variance of the maps
and the expected variance due to instrumental noise: the
1-$\sigma$ confusion noise was found to be 5, 6 and 7 mJy beam$^{-1}$ at 250, 350 and
500 $\mu$m, in agreement with Nguyen et al.\ (2010).
The resulting total 5-$\sigma$ limits are 33, 36, and 45 mJy beam$^{-1}$ respectively.


\begin{figure}
{\psfig{file=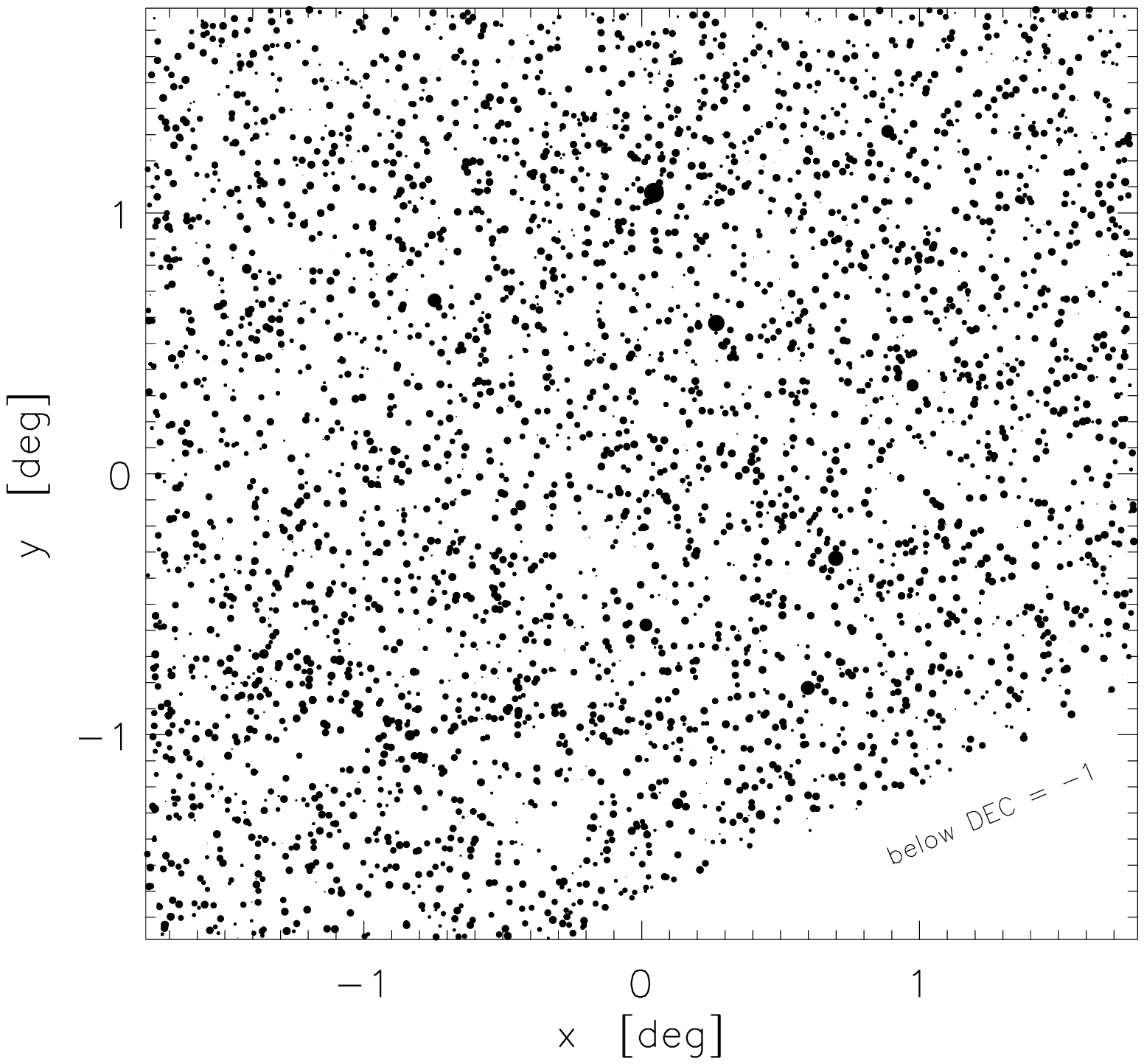,width=8.4cm,silent=1}}
{\psfig{file=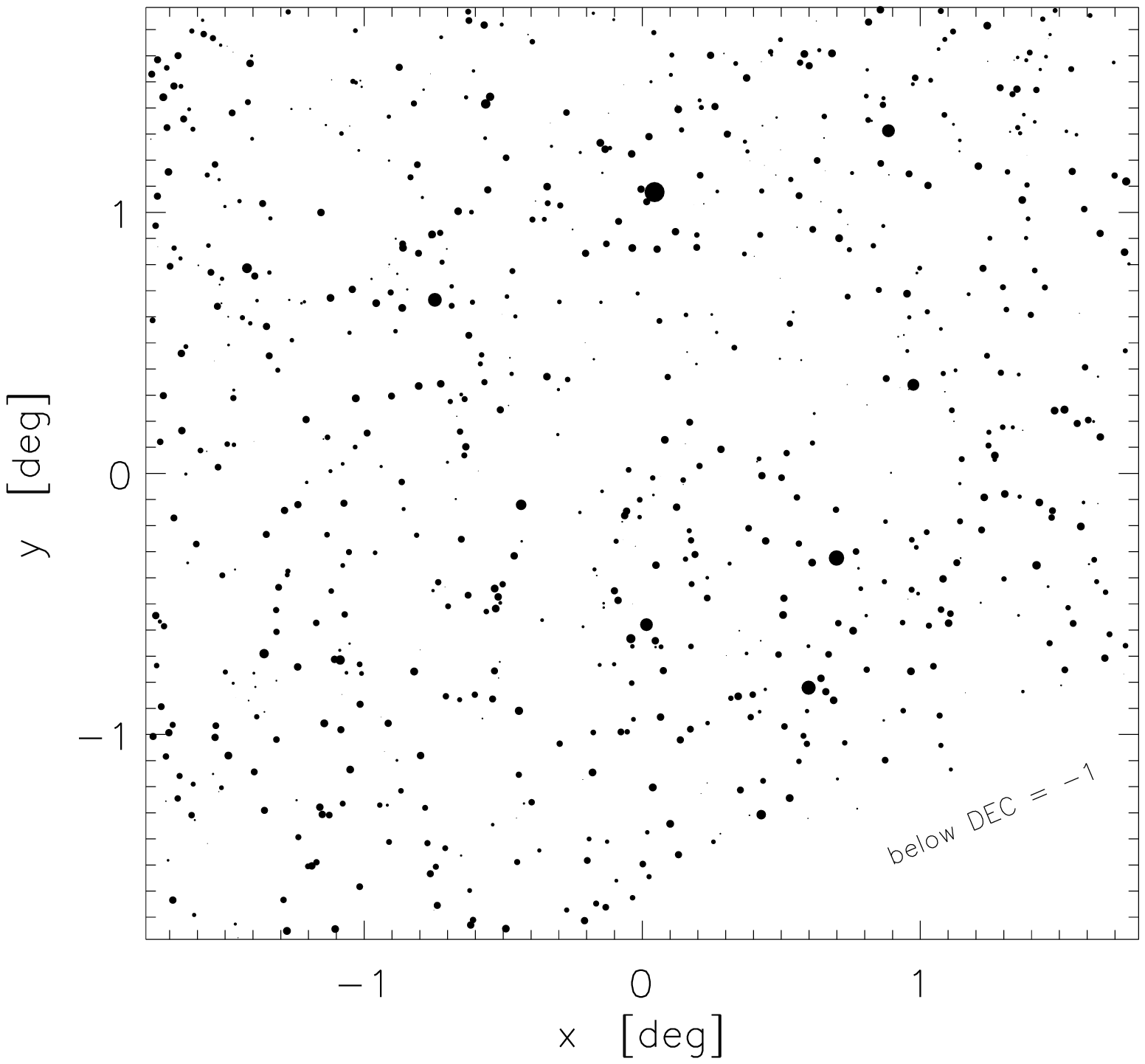,width=8.4cm,silent=1}}
\caption{Top panel: sky distribution of all 250-$\mu$m sources above 33 mJy that
overlap with the GAMA field. These sources cover a wide range of redshifts.
Bottom panel: those sources with spectroscopic redshifts of sufficient
quality (see main text for details), $r_{\rm Pet}<19.4$, and $z<0.3$. 
Both panels were rotated by 25$^{\circ}$ to take into account the orientation of the \herschel\
SDP field, and have no sources in the bottom-right corner which is not covered by GAMA.
Symbol sizes are (inversely) proportional to the sub-mm magnitude at 250 $\mu$m.} 
\end{figure}

\subsection{Redshifts and source selection}


Spectroscopic redshifts are taken from the GAMA database (Driver et al.\ 2009;
Baldry et al.\ 2010; Robotham et al.\ 2010; Hill et al.\ 2011; Driver et al.\ 2011),
which covers most of the {\it Herschel}-ATLAS SDP field, except for a fraction
below declination $\delta = -1^{\circ}$. These redshifts come from a variety of sources,
although most were taken by the GAMA team using 2dF+AAOmega at the AAT
(Driver et al. 2009).

The overlapping area contains 5370 sources for which the SPIRE 250 $\mu$m flux
is above 5$\sigma$ (corresponding to 33~mJy). In order to assign spectroscopic
redshifts from GAMA, which has the SDSS sample as input catalogue, to as many of these sources
as possible, Smith et al.\ (2011)
matched the SPIRE sources to the SDSS DR7 sample (Abazajian et al.\ 2009), using a likelihood-ratio
analysis (Sutherland \& Saunders 1992). All $r<22.4$ SDSS sources within a 10 arcsec radius
of each SPIRE source were considered, taking into account that the true counterpart could be
below the optical magnitude limit. In this process, Smith et al.\ (2011) also
calculated a reliability $R$ for any given source to be the correct counterpart,
and we follow their recommendation to only use sources with $R\ge 0.8$. 

Besides cutting at reliability $R$, we also select GAMA sources with 
redshifts of sufficient quality ($Q\ge$3, see Driver et al.\ 2011 for a detailed
definition of the redshift quality parameter $Q$), and with $r_{\rm Pet}<19.4$, which is the
Petrosian magnitude cut for which GAMA was designed to be complete in redshift coverage;
GAMA has almost achieved that ($98.7\%$ completeness for this cut).

We have tried supplementing these redshifts with (available) photometric redshifts from
Smith et al.\ (2011), selecting only those that have uncertainties $\Delta z < 0.05$,
which is equal to the size of the redshift bins we chose for our redshift slices
(see section 4.1.2). These redshifts are mostly in the range $0.3<z<0.4$, but we found that
their number is too few to give a reliable clustering estimate in this range. Also, the additional
complexity of taking into account the relatively large redshift errors result in clustering
detections that are marginal at best. We therefore chose not to use photometric
redshifts for this study.

Summarizing, we selected low-redshift sub-mm galaxies which:
\itemize{
\item are above 5$\sigma$ at 250 $\mu$m (a flux cut of 33 mJy);
\item are in the GAMA 9h field;
\item have source ID reliability $R\ge0.8$;
\item have spectroscopic redshift quality $Q\ge 3$ 
\item have $r_{Pet}<19.4$;
\item have $z<0.3$.
\medskip
}

The sample thus constructed, with flux limits in both optical and sub-mm bands, leads to a somewhat
specialized selection. In particular at the faint end we select galaxies that are either red, dusty, or both.
For our spectroscopic redshifts we are restricted to $r_{Pet}<19.4$ galaxies, but we do have photometric
redshifts for many galaxies beyond that (Smith et al.\ 2011). This allows us to estimate
the minimum fraction of low-redshift sub-mm galaxies missed in our analysis (i.e.\ those with $r_{Pet}>19.4$
but with a low photometric redshift) for each of the redshift slices considered.
For the \herschel-ATLAS Phase 1 dataset, which is much larger than the SDP dataset considered here,
photometric redshift have recently been estimated (Pearson et al. 2012) for galaxies down to $r_{Pet}\approx 20.8$,
giving sufficient depth and width to give a reasonable incompleteness estimate for our low-redshift sub-mm samples,
even though this is a lower limit (there will still be low-redshift sub-mm galaxies
with $r_{Pet} > 20.8$, although we do not expect this fraction to be large).

Performing this analysis for the G09 Phase 1 field, we find an minimum incompleteness fraction of 14 per cent
for $z<0.3$, where this fraction is very low at the lowest redshift end, and around 40 per cent near $z=0.3$.
These and additional fractions are given and discussed in the relevant sections below.

The top panel of Fig.\ 1 shows all $5\sigma$ SPIRE sources within the GAMA field
(using just the first two criteria that are listed above), 
most of which are not actually identified in the GAMA source catalogue.
Those that are identified as GAMA sources, have $r_{\rm Pet}<19.4$, $z<0.3$,
and are sufficiently reliable matches with good quality redshifts ($R\ge0.8$ and $Q\ge3$),
are shown in the bottom panel of Fig.\ 1. Both panels 
were rotated by 25$^{\circ}$ to take into account the orientation of
the \herschel\ SDP field. The corner region below $\delta = -1^{\circ}$ is the area that is not
part of the GAMA survey, and therefore disregarded for the purpose of this paper.

The redshift distribution of the sources with GAMA redshifts is shown in Fig.\ 2
(all criteria except the last have been applied).
Beyond $z=0.3$ the number of spectroscopic redshifts (thin histogram in Fig.\ 2) quickly decreases,
so we restrict ourselves to $z<0.3$ in this paper.

For the purpose of estimating the spatial clustering length through the Limber equation
inversion method (which involves numerical solutions), we fit the redshift distribution to a
function of the form $n(z)=z^{1.5} e^{a-bz^2}$. The best fit to our observed distribution is for $a=7.0$
and $b=26$     
(shown as a smooth solid line in Fig.\ 2). 


\begin{figure}
{\epsfig{file=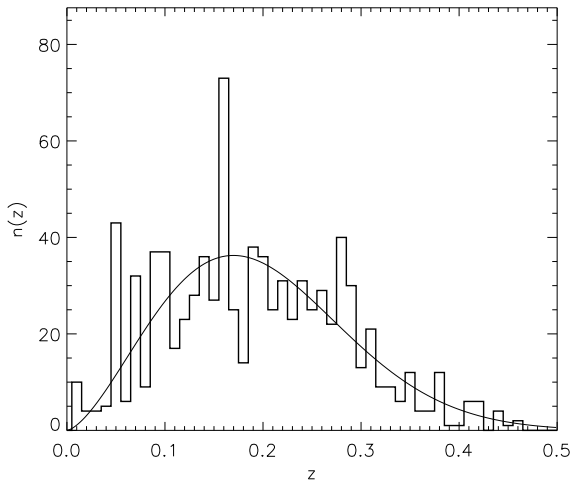,width=8.4cm,silent=1}}
\caption{Redshift distribution of our sample of 250-$\mu$m sources above 33 mJy, with the reliability,
redshift quality, and Petrosian $r$-band cuts applied (see main text for details).
The smooth solid line is a fit to the distribution of all selected redshifts: $n(z)= z^{1.5} e^{7.0-26z^2}$. } 
\end{figure}

\begin{table*}
\centering
  \begin{tabular}{lrrrrr}
  \hline
  \noalign{\vfilneg\vskip -0.2cm}
  \hline
Slice             &     $N$    &  minimum        &       $A$      & $\delta$\hskip 0.5cm &    $r_0$           \\
                  &            &  incompleteness &  [arcmin]      &                      &    [Mpc]           \\
\hline
     all $z          $ &  5363 &       -         & 0.006 $\pm$ 0.008 &  0.51 $\pm$  0.09 &  -                 \\
$     z < 0.3        $ &   724 &  14\%           &  0.28 $\pm$  0.33 &  1.10 $\pm$  0.39 &  6.34 $\pm$  5.44  \\
$     0.05 < z < 0.10$ &   123 &  1.1\%          &  1.09 $\pm$  0.97 &  0.80 $\pm$  0.29 &  3.23 $\pm$  2.19  \\
$     0.10 < z < 0.15$ &   137 &  2.3\%          &  2.45 $\pm$  1.20 &  0.62 $\pm$  0.15 &  4.47 $\pm$  1.72  \\
$     0.15 < z < 0.20$ &   167 &  7\%            &  2.13 $\pm$  0.62 &  0.95 $\pm$  0.19 &  7.20 $\pm$  1.81  \\
$     0.20 < z < 0.25$ &   136 &  20\%           &  0.59 $\pm$  0.66 &  0.58 $\pm$  0.19 &  3.02 $\pm$  2.50  \\
$     0.25 < z < 0.30$ &   145 &  39\%           &  0.66 $\pm$  0.86 &  1.05 $\pm$  0.83 &  5.13 $\pm$  5.50  \\
\hline
\noalign{\vfilneg\vskip -0.2cm}
\hline
\end{tabular}
\caption{Clustering measures for all samples considered (see main text for details) for the
two-parameter fits. An estimate for $r_0$ for the `all z' sample has not been given as its
redshift distribution is unknown.}
\end{table*}

\section{Methods}

\subsection{Estimating the angular correlation function}

The standard estimator for measuring angular correlations is 
$w_{LS}=(DD-2DR+RR)/RR$  (Landy \& Szalay 1993),
where $DD$, $DR$ and $RR$ are the (normalized) galaxy-galaxy,
galaxy-random and random-random pair counts at separation $\theta$.
We employ a more abundant random catalogue (by a factor of ten)
that Poisson samples the same survey region as our observed catalogue. The
normalization takes out the overabundance of the random catalogue by scaling
the $DR$ and $RR$ counts accordingly.
For the estimate of $w(\theta)$ and its errors we use the Jackknife technique
(e.g. Wall \& Jenkins 2003, Norberg et al.\ 2009), employing $4\times4$ regions
and estimating errors from the Jackknife sampling variance.

The estimator is to be fitted by its expected value \\
$1+\langle w_{LS} \rangle = [1+w(\theta)]/(1+w_\Omega)$\ ,
where the `integral constraint' $w_\Omega$ is the integral of the model
for the two-point correlation function over the survey area.
We consider a two-parameter fit for
the generic power-law $w(\theta) = (\theta/A)^{-\delta}$, 
as well as a single parameter fit for $A$ with constant slope $\delta=0.8$.
Van Kampen et al. (2005) give details on the fitting technique, which
employs non-linear $\chi^2$-fitting using the Levenberg-Marquardt method (Press et al.\ 1988).
This allows us to easily take into account the integral constraint,
but also produces the covariance matrix of the fitted parameters which provides
a good estimate of their uncertainties.

For small samples, the estimates depend somewhat on the way the data is binned. For
this reason we have used a range of bin sizes and intervals over which we bin the angular
separations, and adopt the one that produces the smallest fractional uncertainties in
the jackknife estimates. Furthermore, there is some evidence of a feature in the angular
correlation function near the filter scale of 2 arcmin used in the source detection, so we
excluded the corresponding bin from the fit. This was only found to be important for the
full sample (as demonstrated by the fourth datapoint in Figure 3): no difference was
seen for the subsamples introduced in Section 4.2.


\begin{figure}
{\epsfig{file=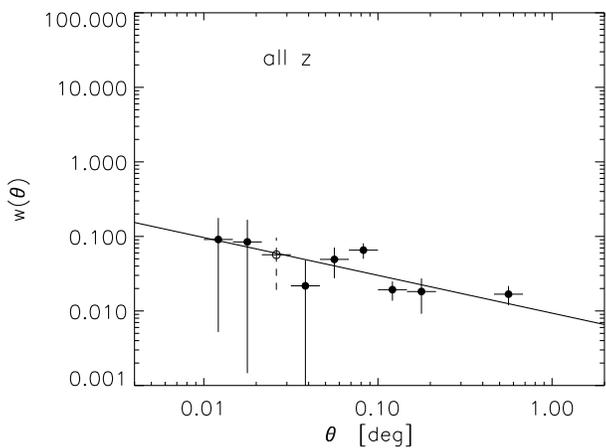,width=8.4cm,silent=}} 
\caption{Angular correlation functions for all 250 $\mu$m galaxies in the SDP field 
(i.e. the sample shown in the top panel of Fig.\ 1).
The solid line shows the two-parameter fit, where the fitted parameter values can be found in the
top row of Table 1. Open symbols represent negative values for the estimated correlation function.
Errors are obtained using the jackknife technique (see main text for details).}
\end{figure}

\subsection{Spatial correlation length from $w(\theta)$}

The traditional method (e.g.\ Peebles 1980) for estimating the spatial clustering length $r_0$
is to measure the angular correlation function and the redshift distribution, and then
use Limber's equation (Limber 1953) to derive $r_0$. We employ the
code used by Farrah et al.\ (2006), assuming a smooth redshift distribution
of the form $n(z) = z^{1.5} e^{a-bz^2}$ that is fitted to the observed redshift distribution.

Because our analysis is limited to low redshift ($z<0.3$), the precise choice for the assumed evolution of the
the correlation function does not matter too much: assuming `stable clustering' (fixed in physical
coordinates) of `comoving clustering' (fixed in comoving coordinates) does not change the results
much for $z<0.3$ (most terms scale with $1+z$), especially because we primarily consider redshift
slices of thickness $\Delta z = 0.05$.

\section{Results}


\subsection{Angular clustering without redshift information}

We first look at the angular correlation function of all sub-mm galaxies detected
by SPIRE at the 5$\sigma$ level, i.e. the sample shown in the top panel of Fig.\ 1.
Most of these galaxies do not have measured redshifts, but because of the strong negative
K-correction at these wavebands we expect a wide range of redshifts (see also Amblard et al.\ 2010).

The resulting angular correlation function, along with a two-parameter power-law fit,
is shown in Fig.\ 3, and displays very little
clustering, as was already shown by Maddox et al. (2010) for the slightly larger full
SDP dataset (that is, including the small region that is not part of the GAMA area). This
should just be seen as a confirmation of the earlier results, and is included here to demonstrate
consistency.

\renewcommand{\thefigure}{\arabic{figure}\alph{subfigure}}
\setcounter{subfigure}{1}

\begin{figure*}
\centering
\begin{tabular}{cc}
\epsfig{file=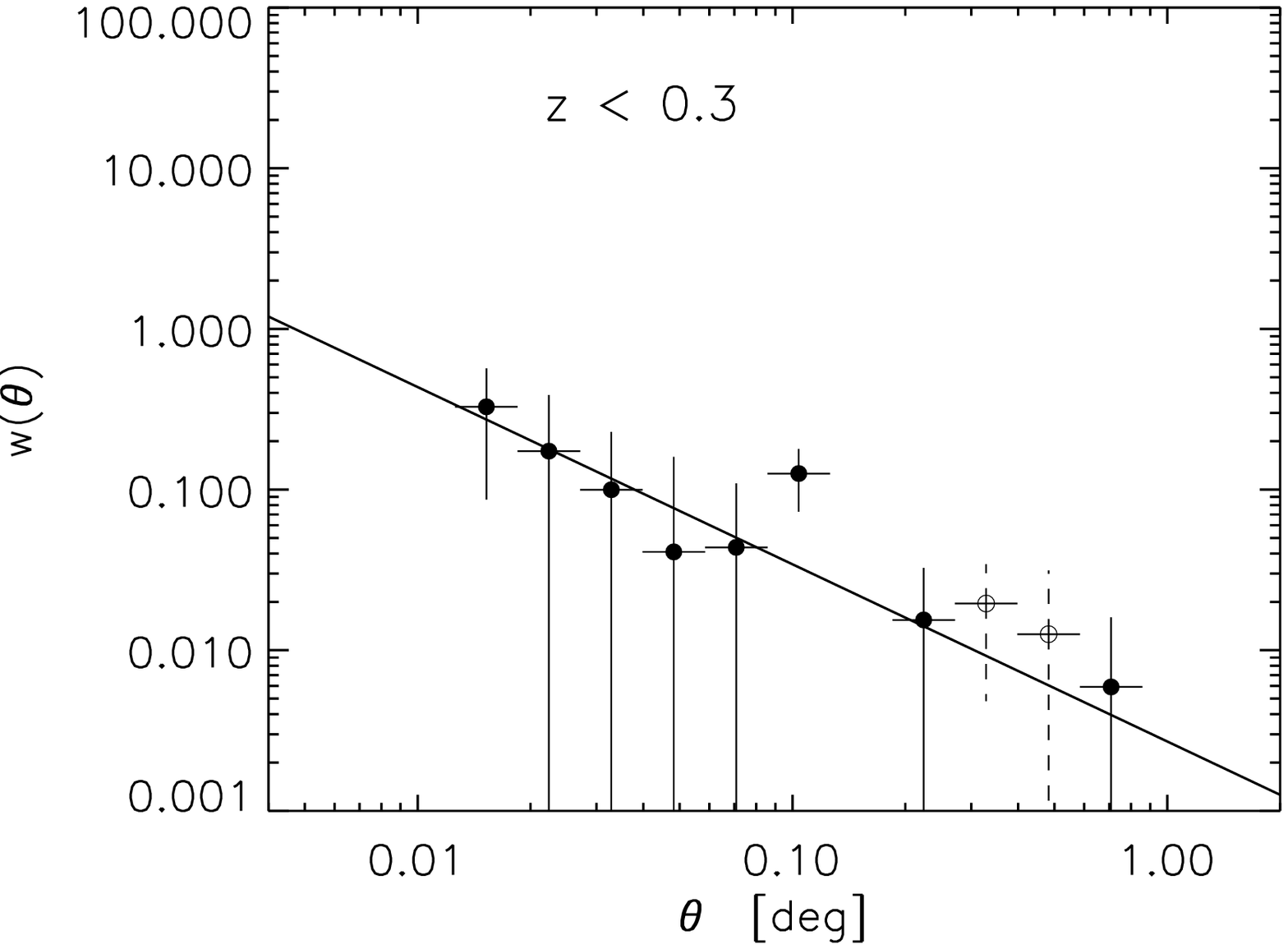,width=0.4\linewidth,clip=} &
\epsfig{file=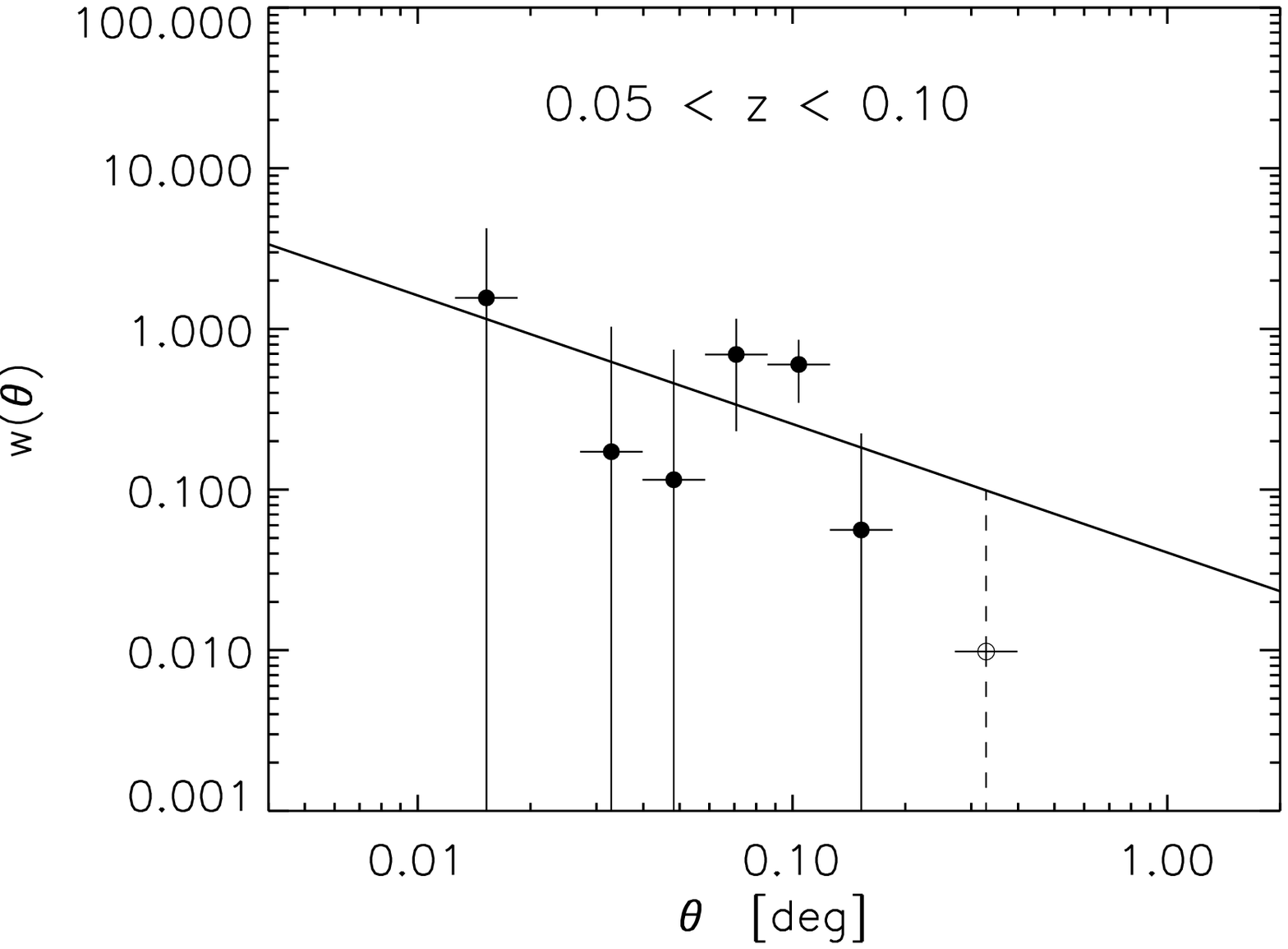,width=0.4\linewidth,clip=}\\
\epsfig{file=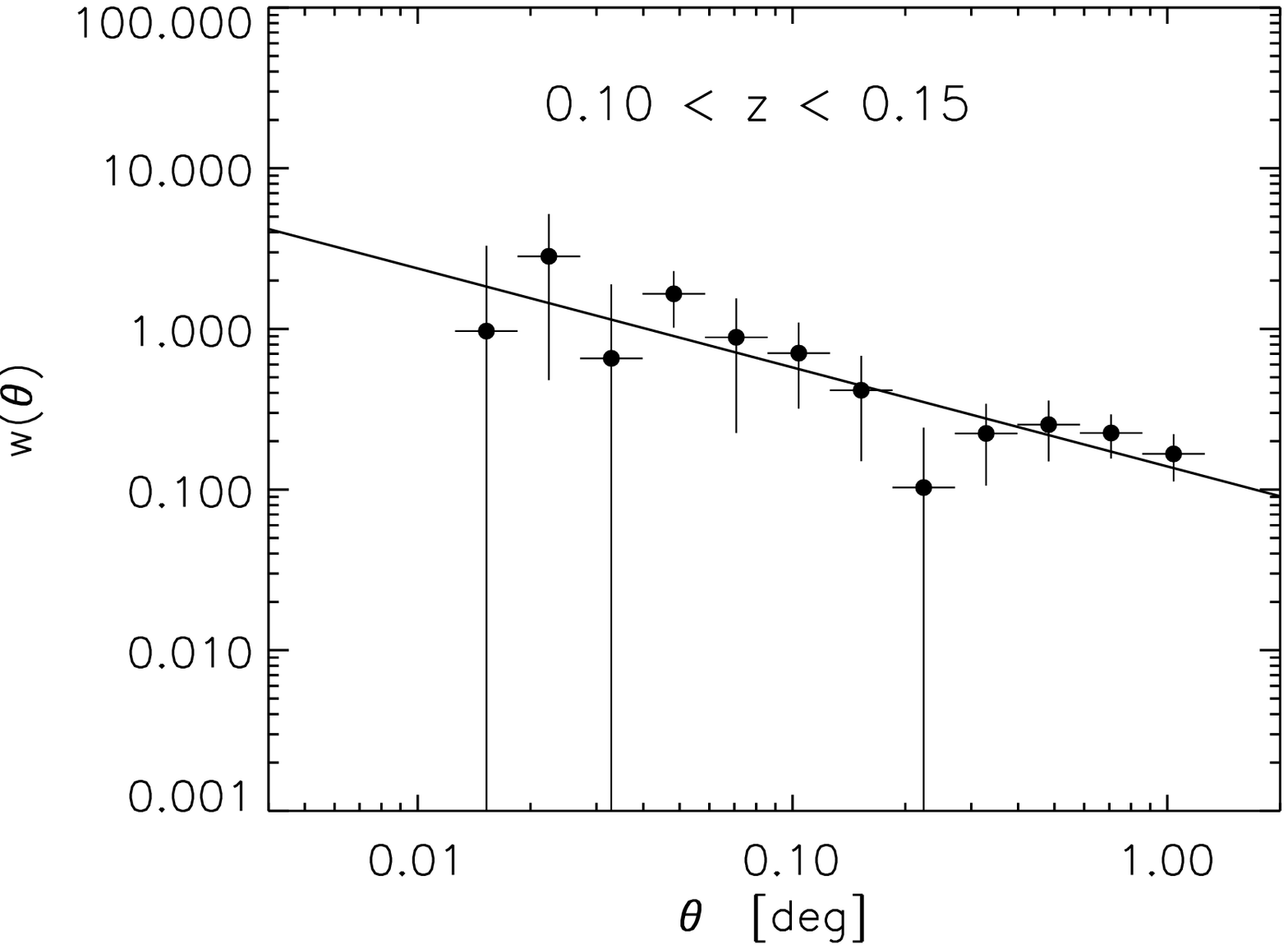,width=0.4\linewidth,clip=} &
\epsfig{file=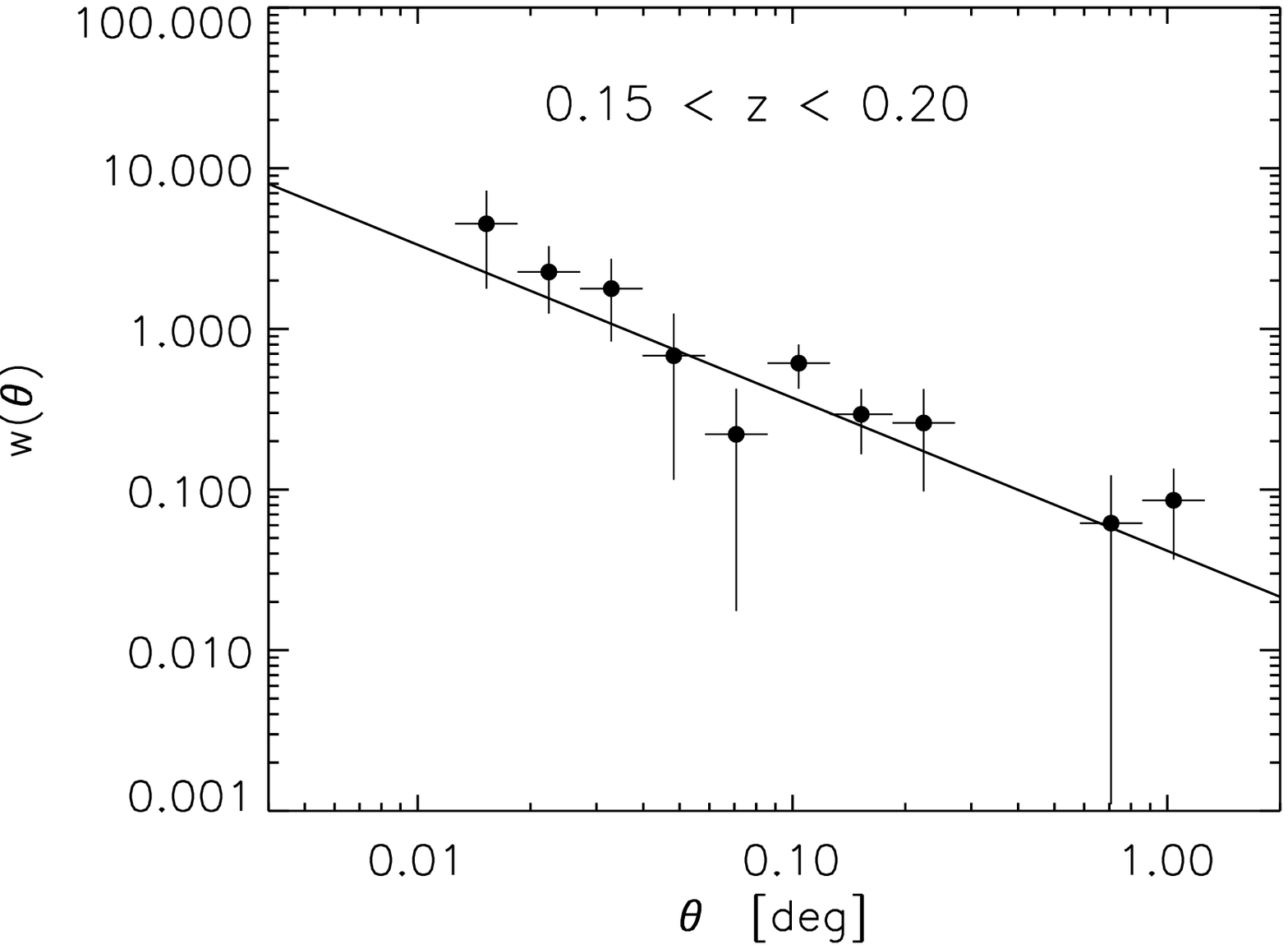,width=0.4\linewidth,clip=}\\
\epsfig{file=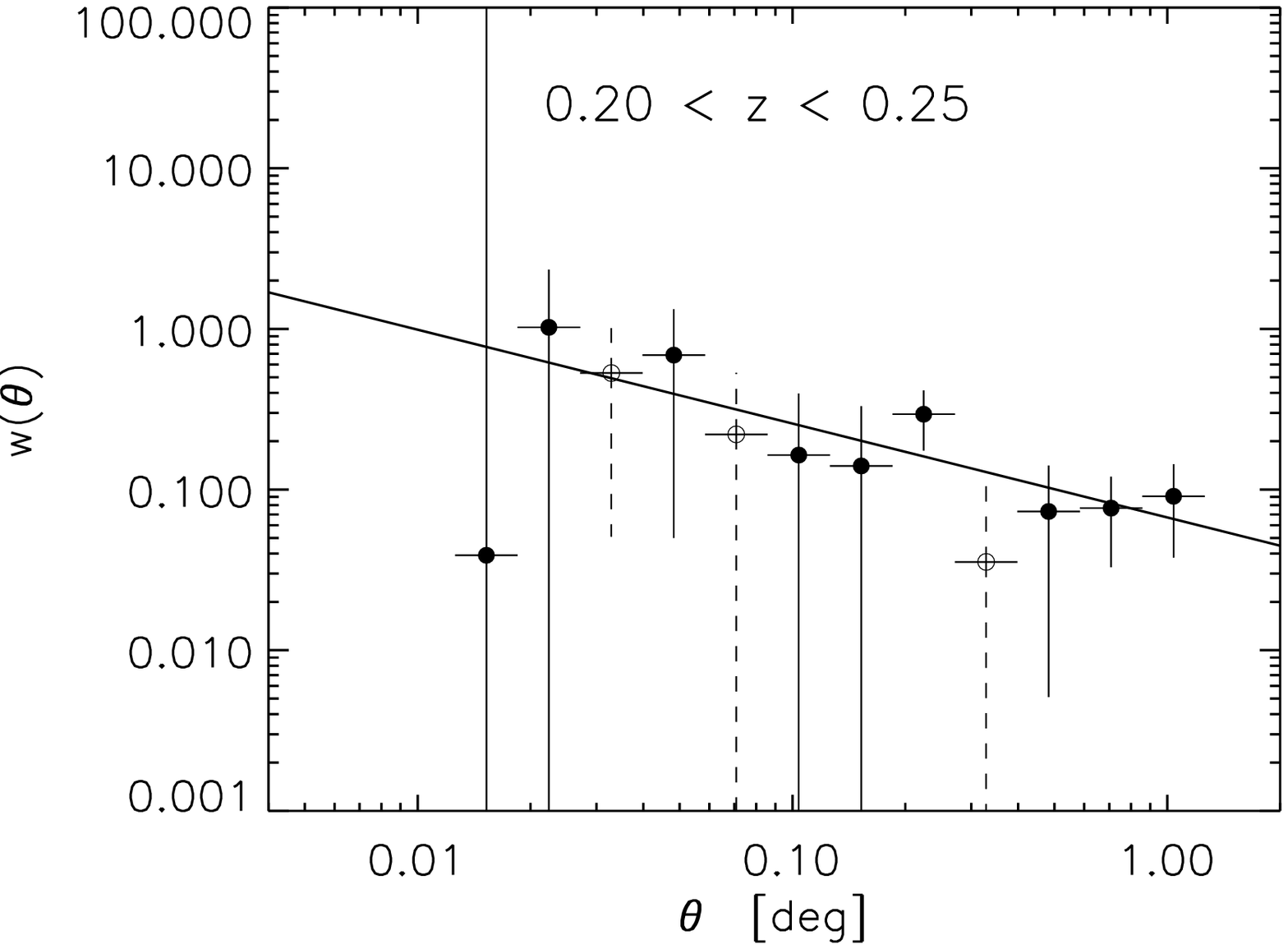,width=0.4\linewidth,clip=} &
\epsfig{file=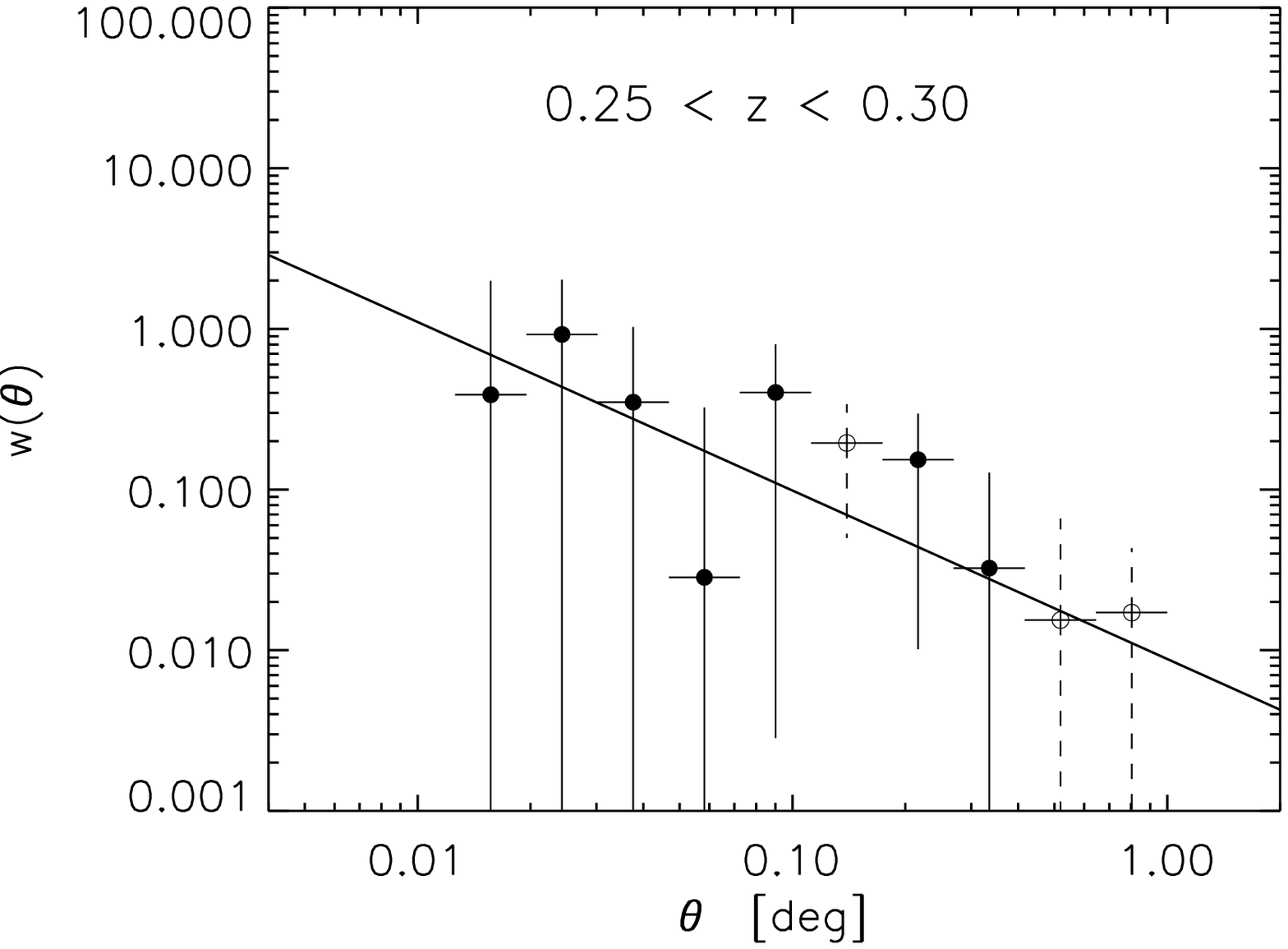,width=0.4\linewidth,clip=} \\
\end{tabular}
\caption{Angular correlation functions for $0<z<0.3$ (top left panel) and for
each redshift slice, with two-parameter fits shown. The values of the fitted
parameters are listed in Table 1. Open symbols represent negative
values for the estimated correlation function. Errors are obtained using the
jackknife technique (see main text for details).} 
\end{figure*}

\addtocounter{figure}{-1}
\addtocounter{subfigure}{1}

\begin{figure*}
\centering
\begin{tabular}{cc}
\epsfig{file=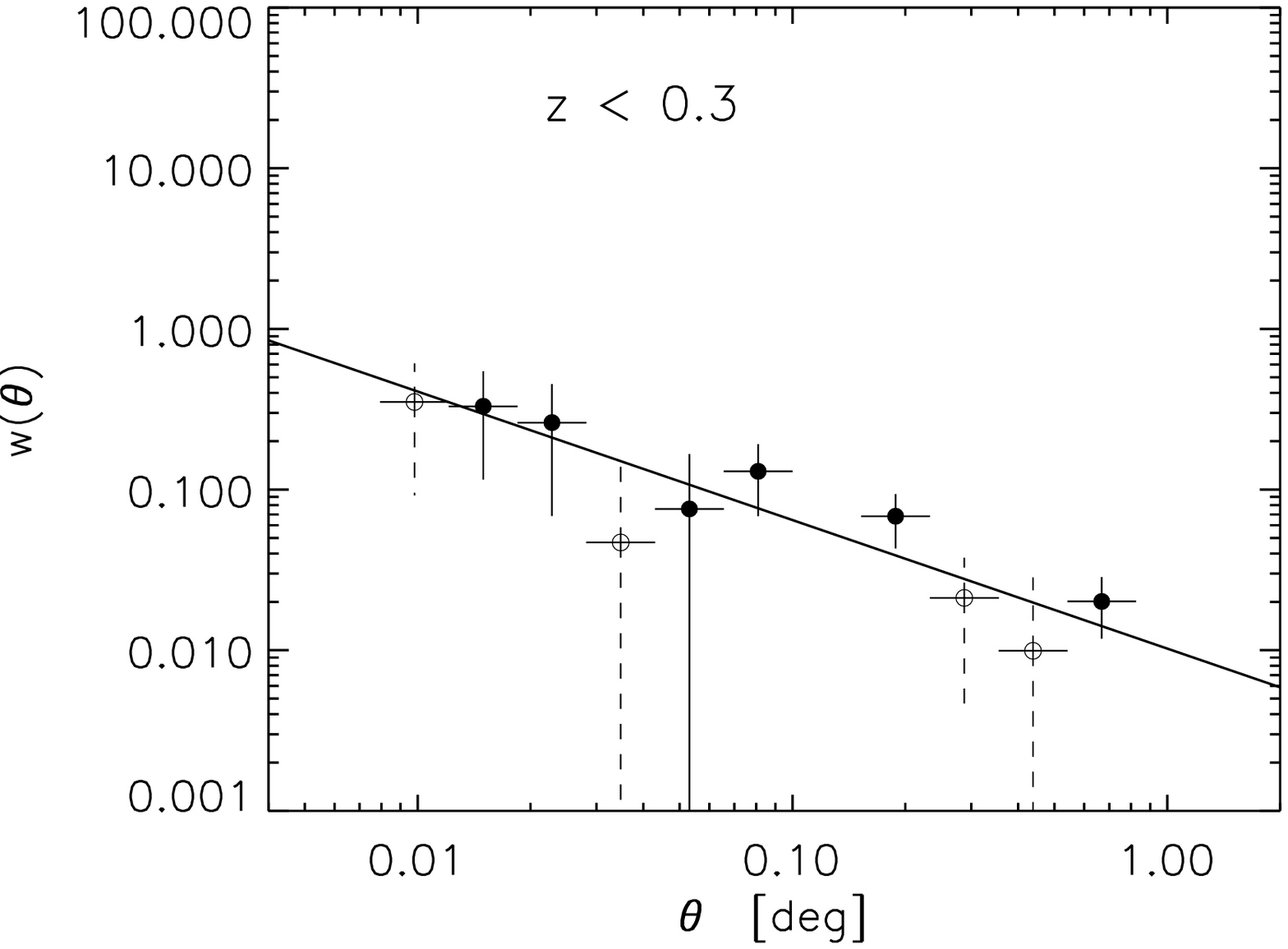,width=0.4\linewidth,clip=} &
\epsfig{file=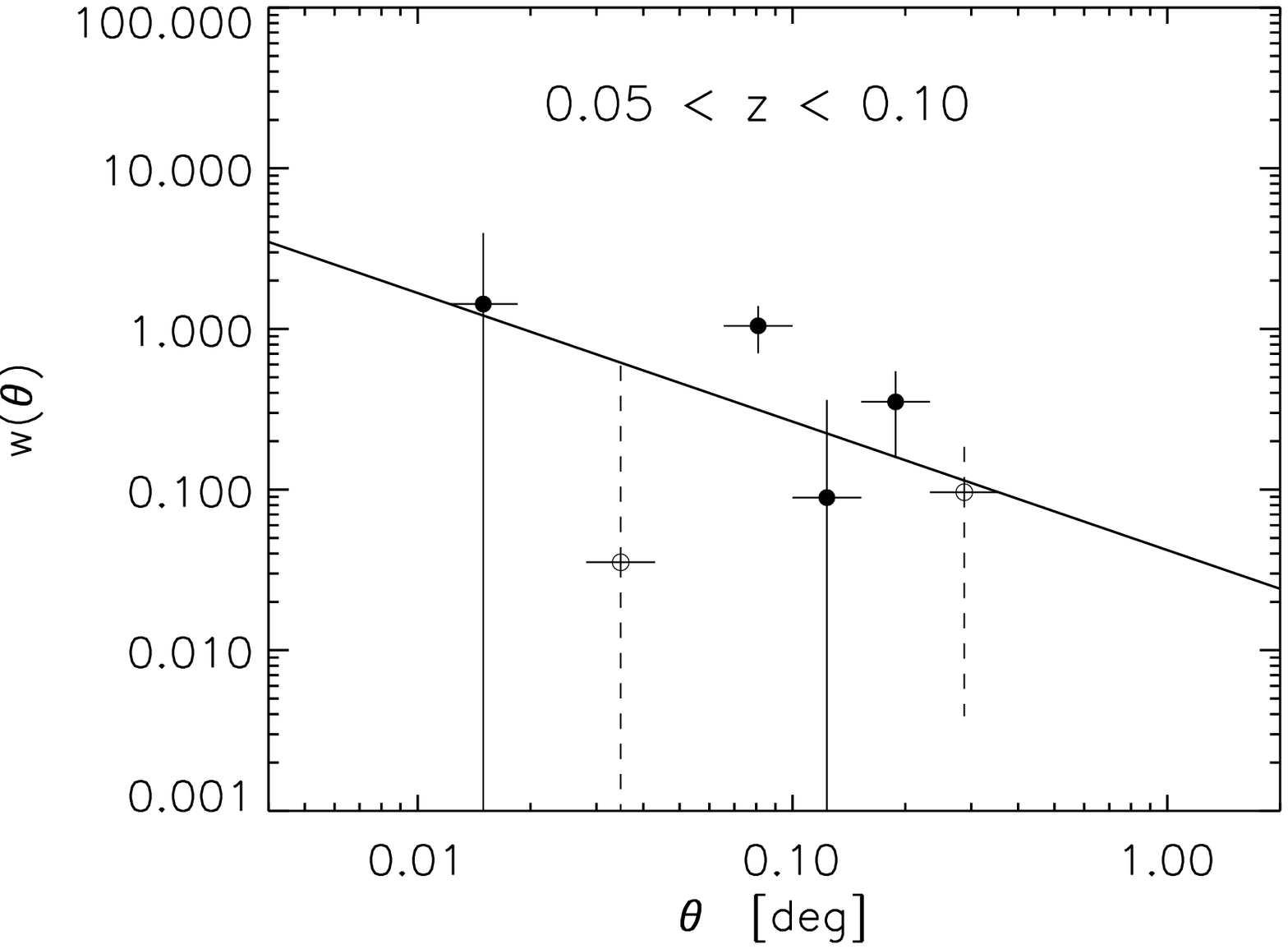,width=0.4\linewidth,clip=}\\
\epsfig{file=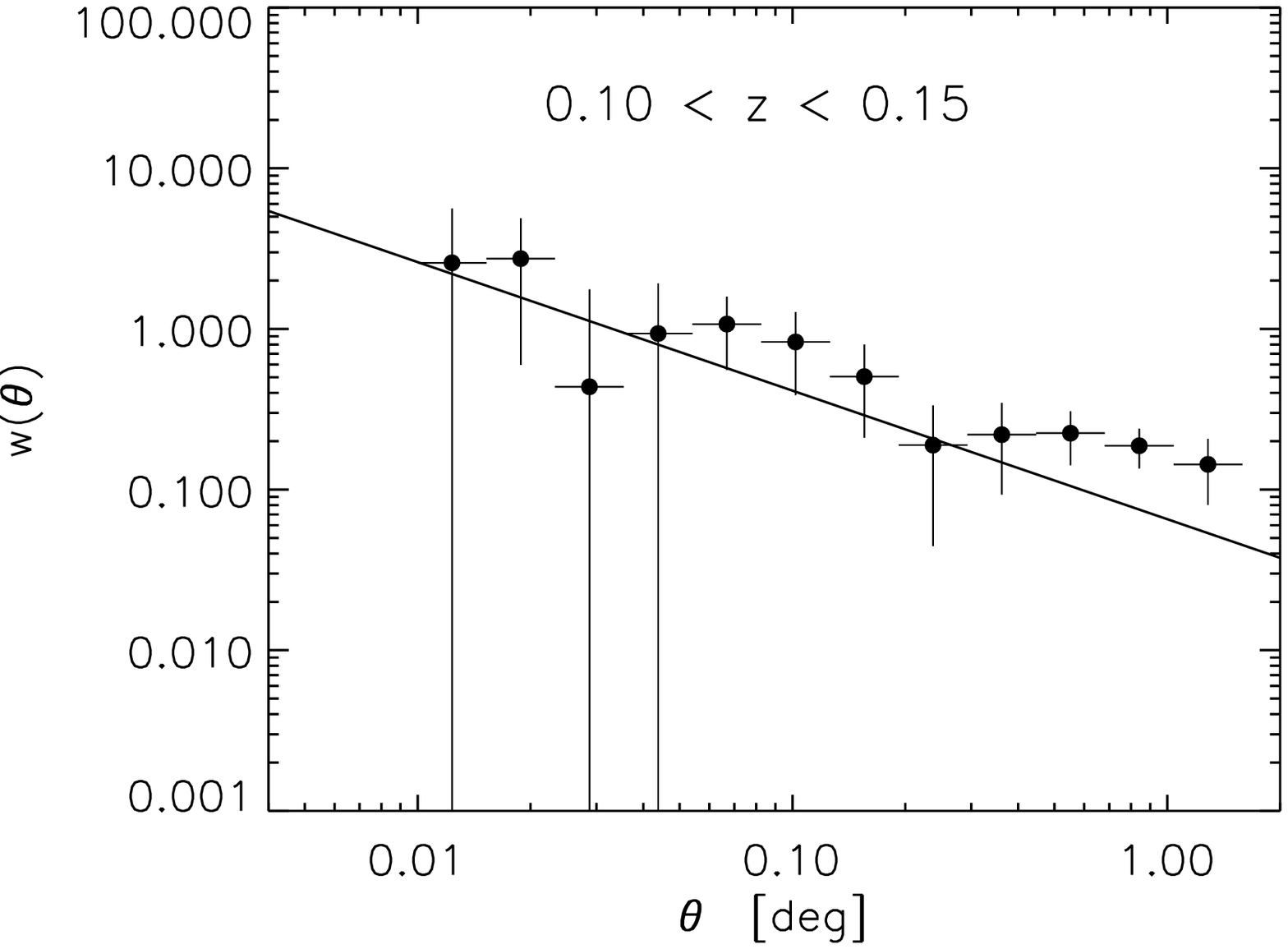,width=0.4\linewidth,clip=} &
\epsfig{file=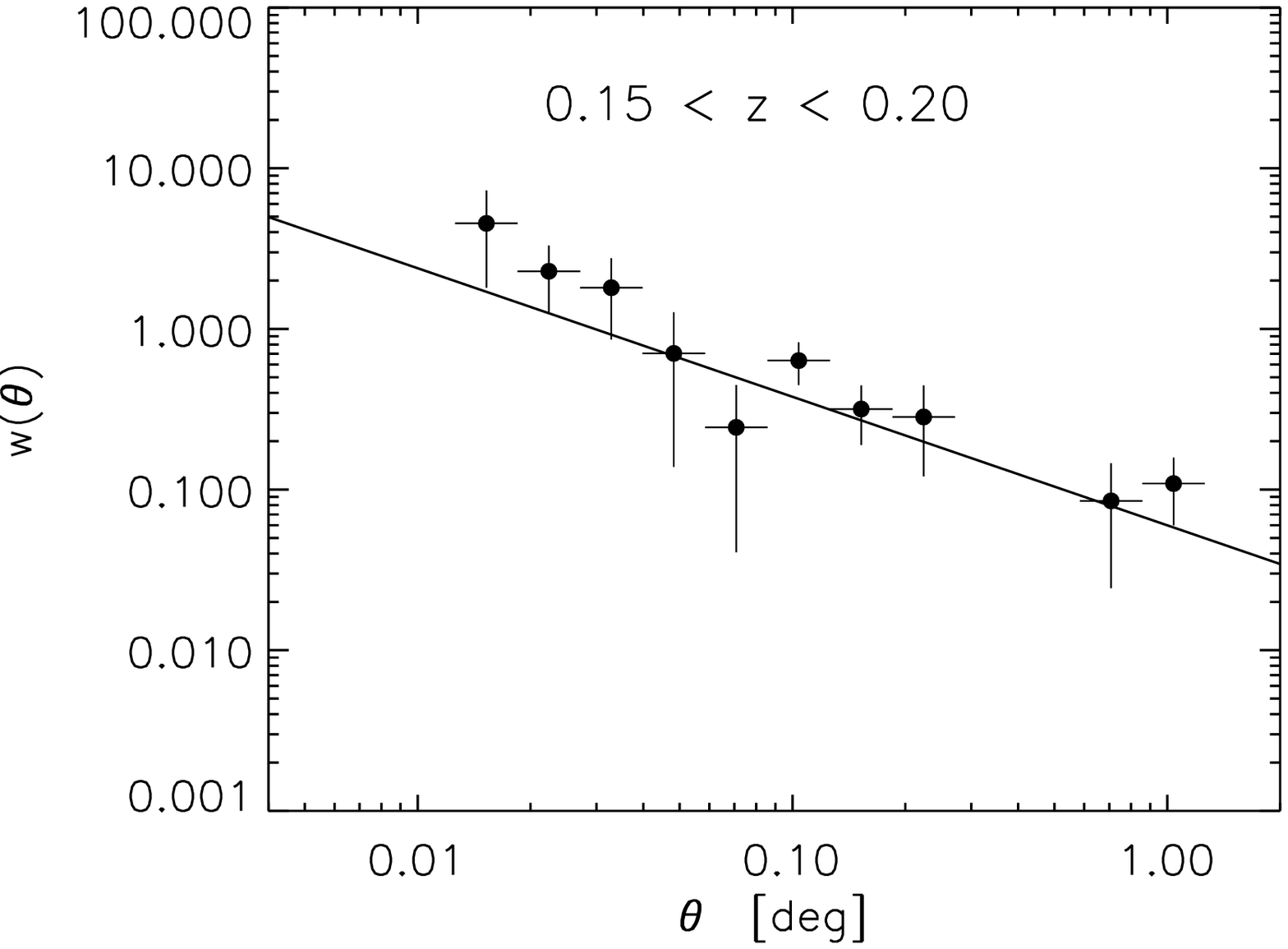,width=0.4\linewidth,clip=}\\
\epsfig{file=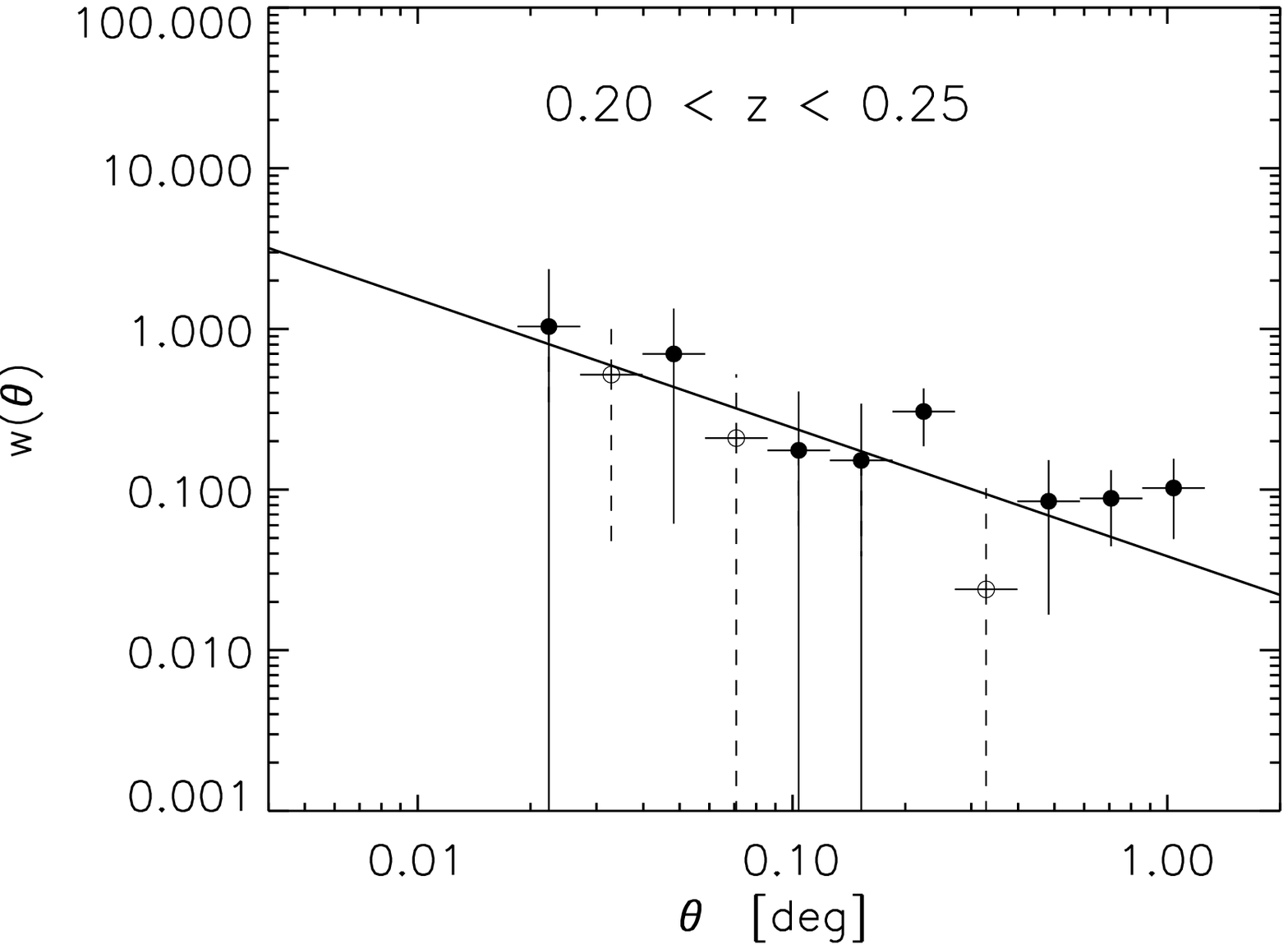,width=0.4\linewidth,clip=} &
\epsfig{file=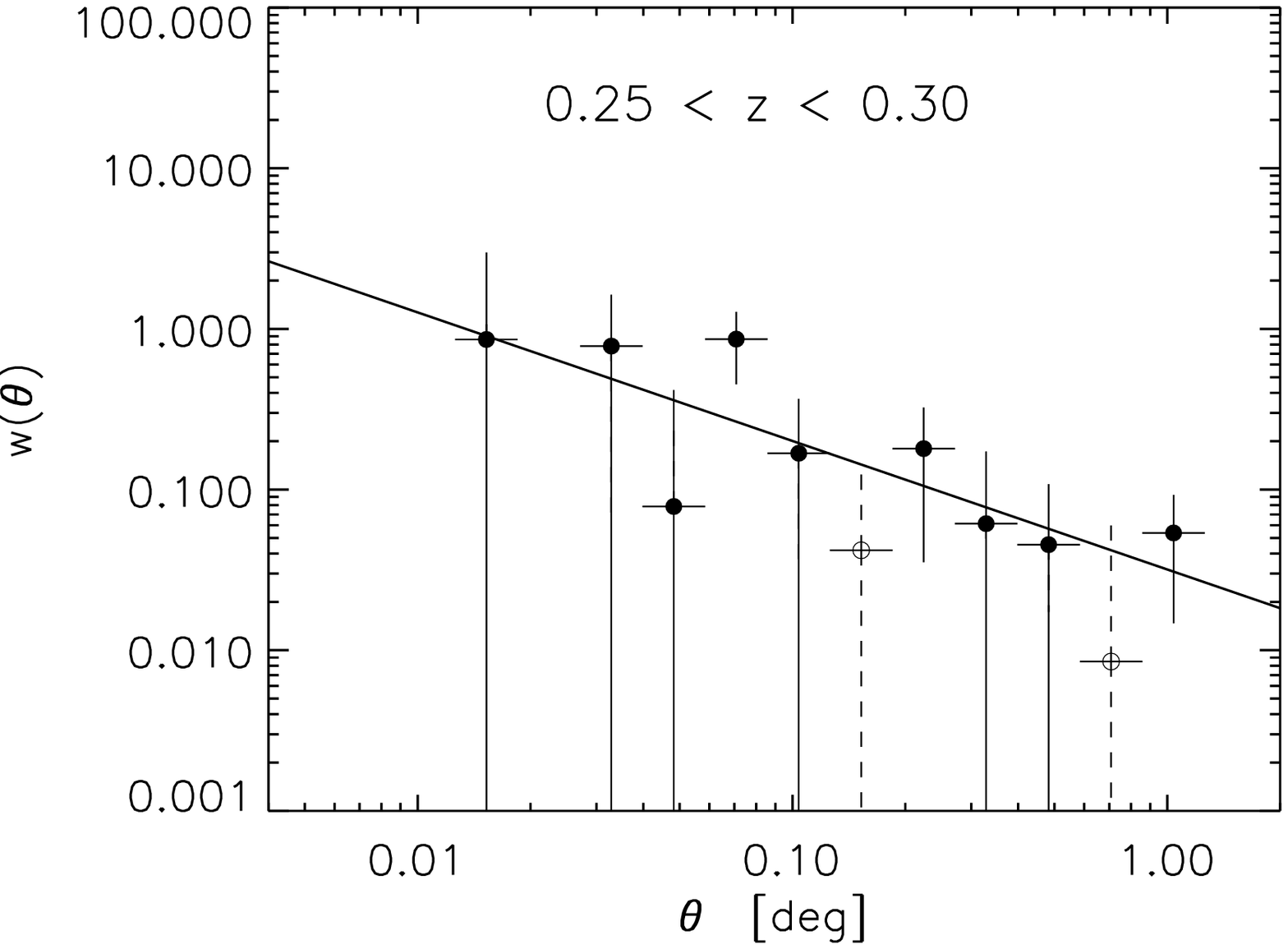,width=0.4\linewidth,clip=} \\
\end{tabular}
\caption{Angular correlation functions for $0<z<0.3$ (top left panel) and for
each redshift slice, with one-parameter fits shown. The values of the fitted
parameters are listed in Table 2. Open symbols represent negative
values for the estimated correlation function. Errors are obtained using the
jackknife technique (see main text for details). Please note that the data points
are not the same as in Fig.\ 4a, as the integral constraint is a function of the
fitting parameters but also used to correct the data points (see main text for 
details).} 
\end{figure*}

\renewcommand{\thefigure}{\arabic{figure}}

\begin{table}
\centering
  \begin{tabular}{lrrr}
  \hline
  \noalign{\vfilneg\vskip -0.2cm}
  \hline
Slice                  &  $N$  &      $A_{0.8}$    &    $r_0$    \\
                       &       &      [arcmin]     &    [Mpc]    \\
\hline
     all $z          $ &  5363 &  0.04 $\pm$  0.02 &  - \\
$     z < 0.3        $ &   724 &  0.20 $\pm$  0.04 &  5.62 $\pm$  1.14 \\
$     0.05 < z < 0.10$ &   123 &  1.14 $\pm$  0.38 &  3.29 $\pm$  1.10 \\
$     0.10 < z < 0.15$ &   137 &  1.99 $\pm$  0.51 &  5.23 $\pm$  1.33 \\
$     0.15 < z < 0.20$ &   167 &  1.78 $\pm$  0.31 &  5.68 $\pm$  1.00 \\
$     0.20 < z < 0.25$ &   136 &  1.20 $\pm$  0.30 &  5.21 $\pm$  1.31 \\
$     0.25 < z < 0.30$ &   145 &  1.13 $\pm$  0.26 &  5.42 $\pm$  1.26 \\
\hline
\noalign{\vfilneg\vskip -0.2cm}
\hline
\end{tabular}
\caption{Clustering measures for all samples considered (see main text for details) for the
one-parameter fits (fixed slope: $\delta=0.8$). An estimate for $r_0$ for the `all z' sample
has not been given as its redshift distribution is unknown.}
\end{table}

\subsection{Angular clustering in redshift slices}

As redshift information is only reasonably complete for $z<0.3$, we now restrict ourselves to
these low redshifts. We first consider the whole range $0<z<0.3$, which is the sample for which
the spatial distribution on the sky is shown in the bottom panel of Fig.\ 1. The SDP field is rotated
by 25$^{\circ}$, so our $x$ and $y$ coordinates represent a rotated coordinate system with respect
to the usual RA and Dec axes.

The angular clustering estimate for these 724 sources is shown in the top left panel of Fig.\ 4a,
along with the two-parameter fit to this estimate. We see that the $z<0.3$ angular clustering signal
is fairly low for the two-parameter fit, with a large uncertainty on the clustering amplitude
(see Table 1): the dilution of any intrinsic
clustering signal due to the long line-of-sight (of order 30 times the width of the lightcone)
is apparently too severe. If we fix the slope at $\delta=0.8$ (as shown in Fig.\ 4b), we find a
more significant one-parameter fit for the amplitude, although the uncertainty is still considerable.

To remove some of the dilution of the clustering signal due to projection along the line-of-sight,
we cut the lightcone in redshift slices. Given the measured redshift distribution, as
shown in Fig.\ 2, we selected redshift slices such that these are as thin as possible but still
contain a sufficient number of sources for a reliable clustering estimate.
We start at $z=0.05$, where the redshift distribution starts to pick up,
and end at $z=0.3$, as the redshift distribution drops sharply there. Our clustering
estimator works well for over a hundred sources, which allows for a redshift
interval of $\Delta z=0.05$ for the present sample (for the full \herschel-ATLAS
data set we should be able to adopt thinner slices). This gives us five redshift slices in all.

The resulting angular clustering estimates are shown in Fig.\ 4a, for the two-parameter fits, and in 
Fig.\ 4b for the one-parameter fits. Please note that the integral constraint (see Section 3.1) is a
function of amplitude $A$ and slope $\delta$, but is also used to correct the data points. This results
in somewhat different data points in Figs.\ 4a and 4b (best seen at the larger angles), as the fitted
values for $A$ and $\delta$ for the two-parameter fit will be different from the fitted value for $A$
and the choice $\delta=0.8$ for the one-parameter fit.

The fitting parameters corresponding to the estimates shown in Figs.\ 4a and 4b are listed in Tables 1 and 2,
for the two- and one-parameter fits, respectively, where $N$ is the number of sources in each slice.
The spatial clustering length $r_0$, as listed in the last column of each table, is discussed
in the following section.

For the one-parameter fit a fixed slope with $\delta=0.8$ was adopted, as found for local, optically selected
galaxies. This value is also consistent with all fitted slopes found from the two-parameter fit, taking
the uncertainties into account. It might therefore be argued that adopting a fixed slope $\delta=0.8$ is
a good approximation and helps to produce a tighter constraint on the clustering amplitude. Still, for both the two- and
one-parameter models we obtain a good fit for each of the redshift slices, even though the angular correlation
function measure itself can be noisy for some of the redshift slices.
The best determinations are for the bins around the peak of the redshift distribution, where we find the most
significant values for the fitting parameters. The slope is in all cases consistent with that for normal galaxies,
i.e. around $\delta\approx 0.8$, even though uncertainties can be considerable, and certainly for the higher
redshift bins is somewhat self-induced due to the optical selection (see Section 2.2) and the choice $\delta=0.8$
for the fixed slope in the case of the one-parameter fits. This is discussed further in Section 5.

\subsection{Spatial clustering}

In order to use the Limber equation inversion to estimate the spatial clustering strength
we need to use the redshift distribution, as shown in Fig.\ 2. Because of the relatively small
field of view, this distribution shows quite a bit of variance. To ensure that the necessary numerical
integrations are well behaved, we fit a continuous function of the form $z^{1.5} e^{a-bz^2}$ to the observed
redshift distribution, which is shown in Fig.\ 2 as a smooth solid line. For the fitting parameters
we found $a=7.0$ and $b=26$. The redshift distribution for all sources is unknown at this point,
so we do not estimate a spatial clustering length for the full sample (first row in Table 1).

We then compute the spatial correlation length $r_0$ for each redshift slice using the Limber
equation inversion technique (see section 3.2), for both the two- and one-parameter fits, assuming
comoving clustering (if we assume stable clustering we obtain values for $r_0$ which are roughly
10 per cent larger). For the spectroscopic redshifts we need not worry much about the redshift errors,
as these are much smaller than the thickness of the slices (i.e. $\Delta z = 0.05$). 

The resulting values for $r_0$ are listed in Tables 1 and 2, respectively, in units of Mpc.
As the resulting value is sensitive to the slope of the angular correlation function, we obtain
different results for the two- and one-parameter fits when the fitted slope (for the two-parameter fit)
is different from 0.8.
Because of the significant uncertainties in the estimates of the clustering lengths, it is premature
to study the evolution of clustering with this data set.
What is interesting is that the slices are sufficiently thin to clearly detect clustering,
which is not the case for the larger redshift range $0<z<0.3$. Once the whole area
of \herschel-ATLAS is covered, we should be able to reduce uncertainties to such a level where
we can systematically study clustering trends with redshift.

\subsection{A structure at $z$=0.164}


\begin{figure}
{\psfig{file=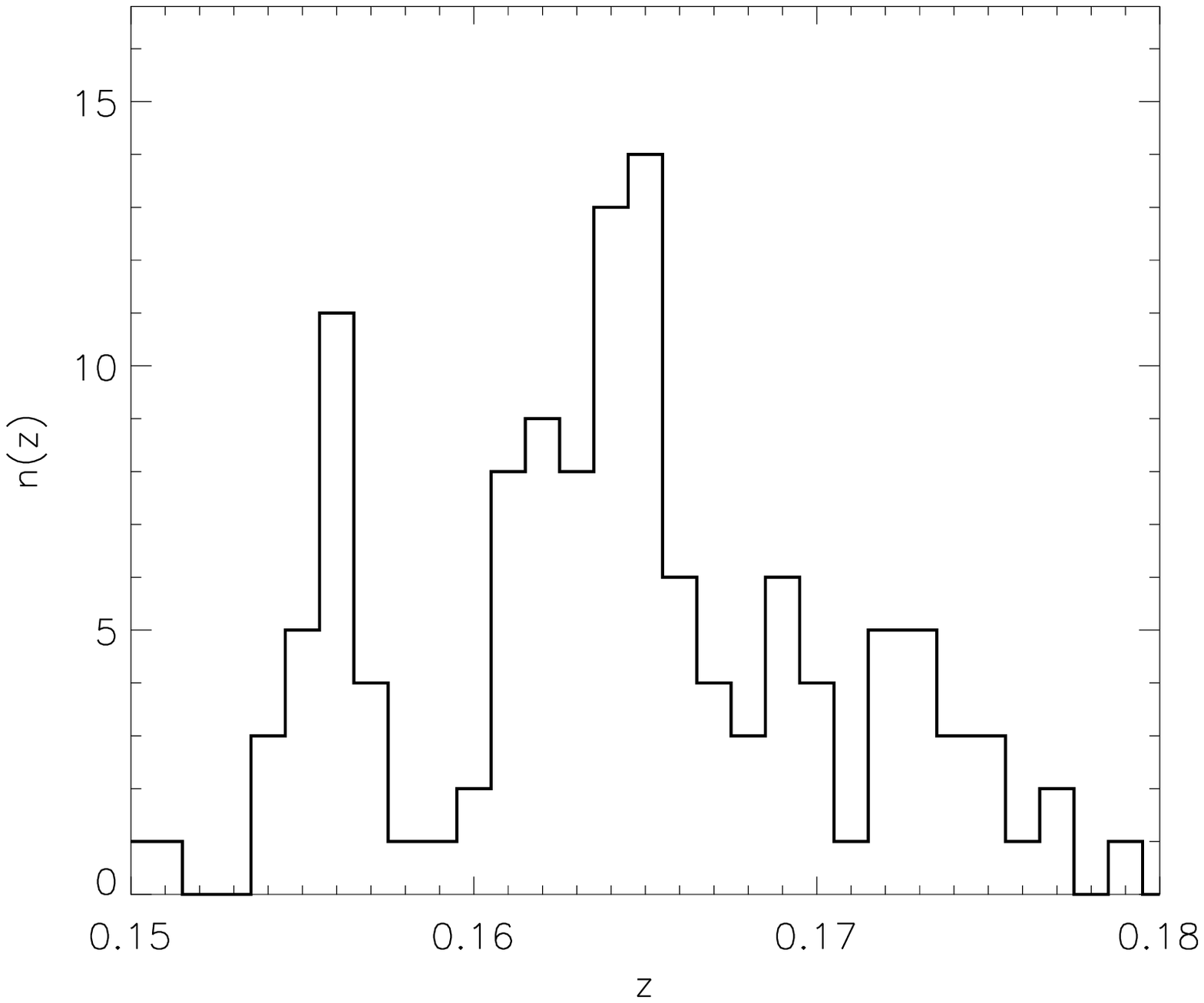,width=8.1cm,silent=1}}
\vskip -0.1cm
{\psfig{file=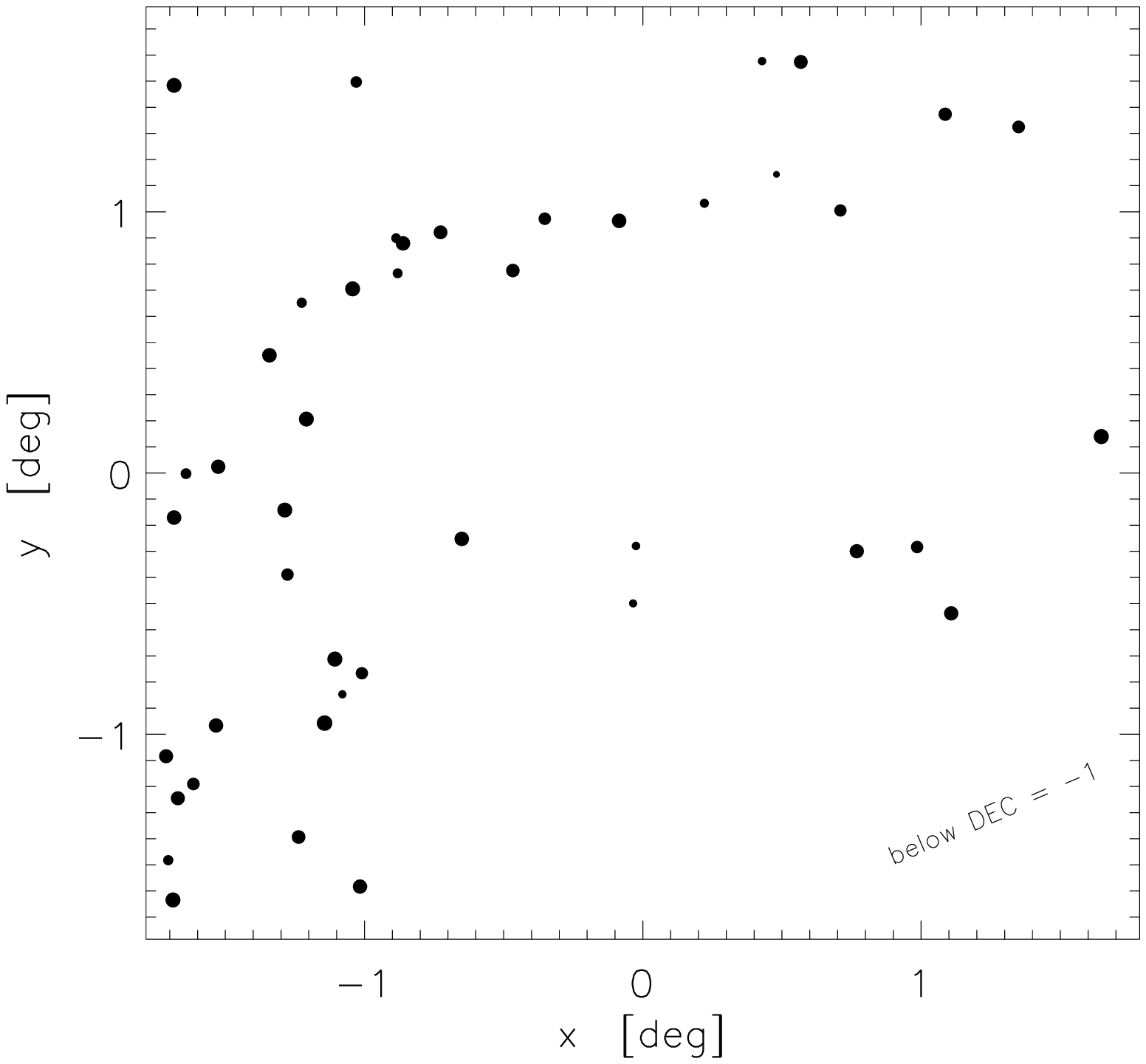,width=8.2cm,silent=1}}
\vskip -0.1cm
{\psfig{file=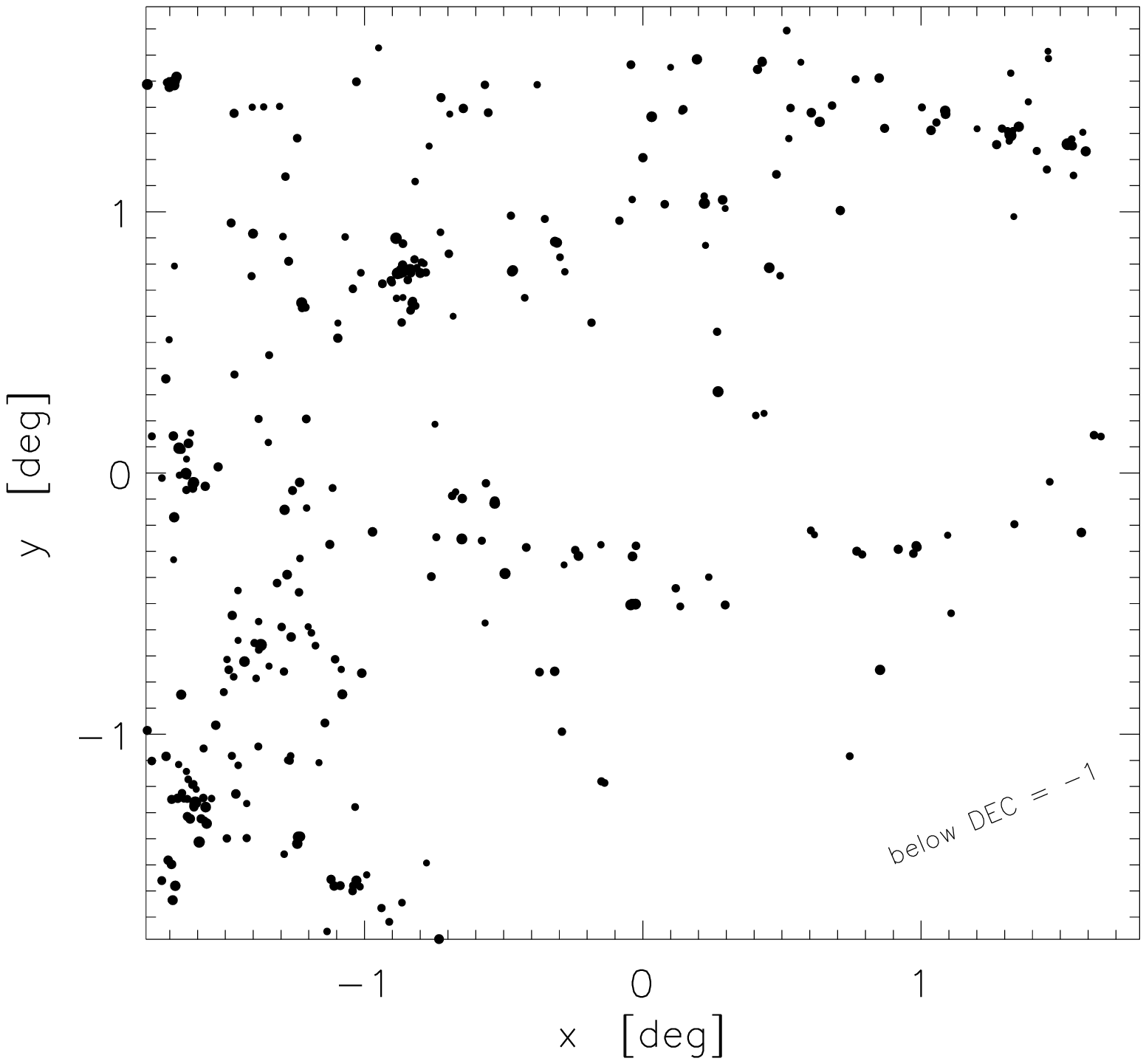,width=8.2cm,silent=1}}
\caption{A structure at z=0.164. The top panel shows a zoom-in on the redshift distribution around
the spike that is prominently visible in Fig.\ 2. Sub-mm sources in the redshift slice $0.162<z<0.166$,
with all other selection criteria applied (see Section 2.2), are shown in the middle panel. For clarity,
symbol sizes are three times larger than for Fig.\ 1. The bottom panel displays all $r_{\rm Pet}<19.4$
GAMA galaxies with redshifts in the same $0.162<z<0.166$ slice.} 
\end{figure}

\begin{figure}
{\psfig{file=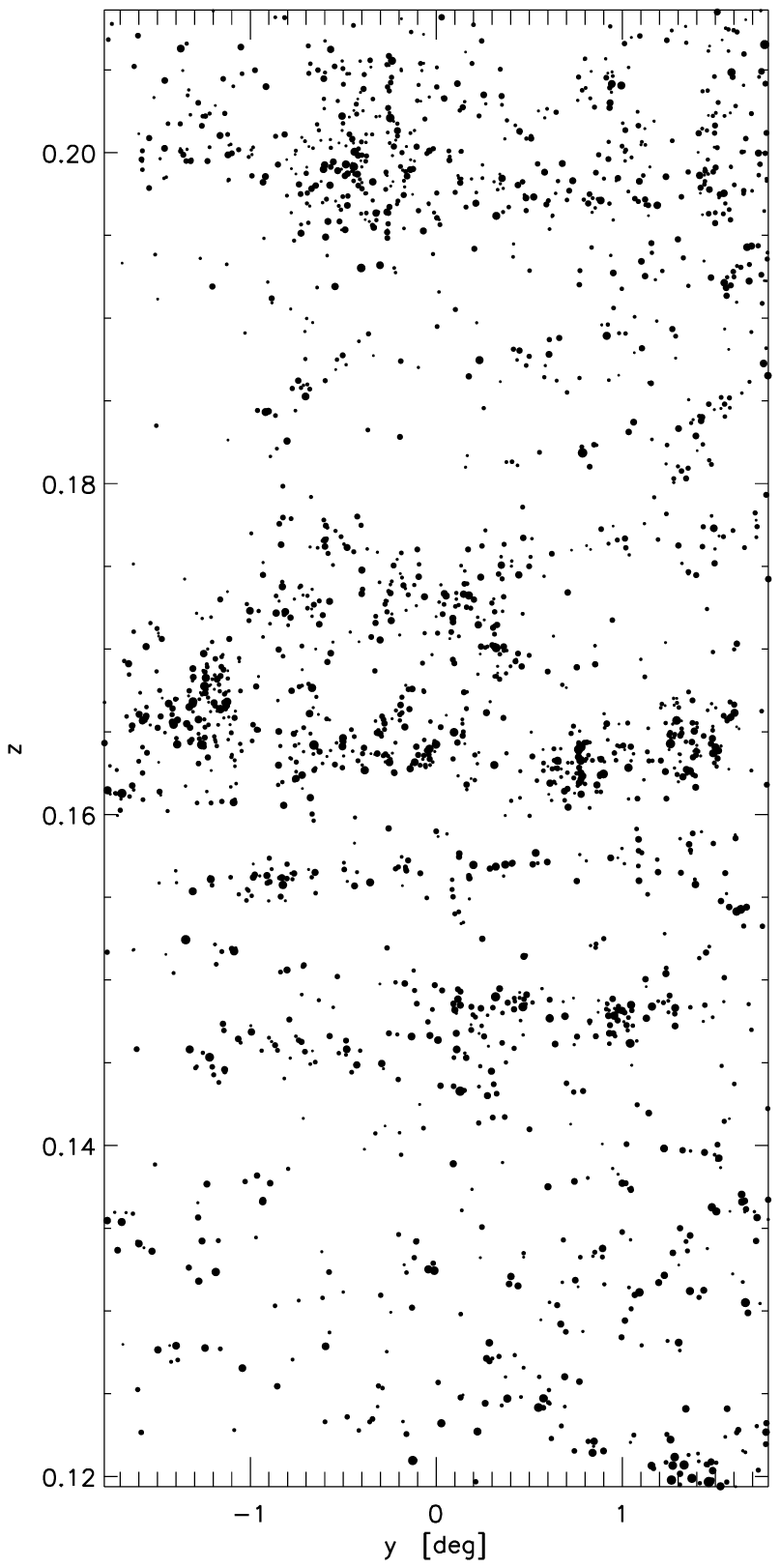,width=8.0cm,silent=1}}
\caption{The z=0.164 structure along the line-of-sight seen in GAMA for all galaxies with redshifts
and $r_{\rm Pet}<19.4$. The `y' coordinate is the same as used in Figs.\ 1 and 5, i.e. one of the
coordinates of the SDP field that is rotated by 25 degrees with respect to RA and Dec. } 
\end{figure}

The redshift distribution of our selected source population, as shown in Fig.\ 2,
displays a fairly pronounced peak near $z\approx 0.16$. If we zoom in on the redshift
distribution around this peak, as plotted in Fig.\ 5 (top panel), we see that the peak is split in
two, with the largest part centred around $z=0.164$. Plotting the distribution of the corresponding
sources (Fig.\ 5, bottom panel) around this redshift, in the interval $0.162<z<0.166$, we see an
clear structure appearing. The thickness of this redshift slice is around 15 Mpc,
whereas at this redshift the field-of-view is around 34 Mpc, so we are likely seeing
filamentary structure mostly aligned with the plane of the sky.

Maddox et al.\ (2010), who performed the first angular clustering analysis of this field, pointed out
that there are patchy wisps of cirrus that plague the SDP field. However, this does not seem to align
with the `filamentary structure' shown in Fig.\ 5. Comparing to the distribution of galaxies
selected in the optical, that is, all GAMA galaxies with redshifts $0.162<z<0.166$ and $r_{\rm Pet}<19.4$
(bottom panel of Fig.\ 5), we see a similar filamentary structure, tracing the low-redshift sub-mm galaxy distribution
fairly well, but also displaying several distinct galaxy groups which are absent in the low-redshift sub-mm
galaxy distribution. 
In Fig.\ 6 we plot this same population of GAMA galaxies along the line of sight, which clearly
shows the overdensity of galaxies at $z=0.164$ to be produced by several groups and filaments roughly
perpendicular to the line-of-sight.
The $z=0.164$ structure might be responsible for the somewhat larger clustering length seen in the $0.15<z<0.2$ 
redshift bin (see Figure 4 and Tables 1 and 2), although the excess is marginally significant.  

\section{Conclusions}

We selected a sample of 250-$\mu$m sources, detected at the 5$\sigma$ level,
from the \herschel-ATLAS SDP data, and used the cross-match
with the GAMA catalogue to assign redshifts to as many of these sources as possible,
taking care to only use reliable IDs and spectroscopic redshifts of sufficient quality,
and exploiting the near completeness of the GAMA spectroscopic redshifts for $r_{\rm Pet}<19.4$.
Because the redshift distribution drops off fairly quickly beyond 
$z\approx 0.3$, we restricted our analysis to $z<0.3$.

Simply taking all 250-$\mu$m sources and ignoring any redshift information reproduces the result
of Maddox et al.\ (2010) for the same data set: no clear clustering signal for these $\sim$5000 sources.
Taking just the 910 galaxies at $z<0.3$ that satisfy our selection criteria, we do detect clustering.
When we subdivide these galaxies into redshift slices of thickness $\Delta z=0.05$ we detect a 
significant clustering signal for most of the slices. 
Clearly the dilution of the clustering signal along the line of sight is too strong
for the relatively small area we have covered for the SDP data, but the results for the slices show that
for the full area of \herschel-ATLAS we should be able to cleanly detect a clustering signal, and
study how this evolves with redshift.

In a related paper, Guo et al.\ (2011) measured the cross-correlation function of the \herschel-ATLAS SDP
and GAMA sources, but also their autocorrelation functions. They used a different method, but
for a similar sample.
Their estimate for the spatial clustering length of $4.76\pm0.63$ Mpc, obtained for a redshift distribution
peaking at $z=0.19$, is consistent with our measures for the (thinner) redshift slices in the same range (both
for the two- and one-parameter fits).
Even though our clustering lengths are somewhat larger in most of our bins, the uncertainties are too large
to establish whether there is a significant discrepancy between our results: we will need the full area of
\herschel-ATLAS to test this.

Our estimates for the five redshift slices (in the range $0.05<z<0.3$) give a mean spatial clustering length
of $4.6\pm 3.4$ Mpc for the two-parameter fits, and $5.0\pm 1.2$ Mpc for the one-parameter fits
(the smaller error is due to the assumption for the slope of the correlation function).
These values are very close to what is found for optically selected galaxies in the SDSS: Zedavi et al. (2011)
found spatial clustering lengths between roughly 4 and 10 Mpc, depending on the absolute luminosity threshold adopted,
with $r_0$ increasing for increasingly brighter thresholds. Most of their subsamples have $r_0$ in the range 5--6 Mpc.
For our special selection criteria we have a distribution over optical luminosities that is
not sharply cut-off, so a straightforward comparison cannot be made, although our spatial clustering lengths are
in the same range. From this point of view, our $z<0.3$ 250-$\mu$m selected galaxies do not appear to be very
different from optically selected galaxies, edging towards blue galaxies (e.g. Coil et al. 2008, Zehavi et al. 2011,
Christodoulou 2012), which would indicate that our sample consists mostly of moderately star-forming galaxies.
We will investigate this in more detail for the full \herschel-ATLAS sample.

One caveat in the comparison to optically-selected galaxies is that our samples are not complete
in the sub-mm band due to the optical limit $r_{Pet}<19.4$, which means that we miss low-redshift galaxies
with $r_{Pet}>19.4$ that are still bright enough in the sub-mm to be included in our sample.
In Section 2.2 we introduced and estimated the minimum fraction missed for each of our subsamples (listed in Table 1).
This is a minimum fraction as the photometric redshifts used in this estimate go deeper (to $r_{Pet}\sim 20.8$) but still
do not include all possible $z<0.3$ sub-mm galaxies. For the lowest redshift slices this fraction is small, so the
conclusion that sub-mm galaxies in that redshift range cluster like optical galaxies is fairly robust. However,
for the highest redshift slice ($0.25<z<0.3$) the incompleteness is at least 39 per cent, and the optical selection
starts to become fairly dominant, making the comparison to a fully optically-selected sample somewhat self-induced.

A more general caveat is that any incompleteness in the sample could bias the clustering estimate if any of our
(sub)samples is not fair, i.e. not a good representation of the sub-mm (sub)sample that would include all
$r_{Pet}>19.4$ sources as well.
As the incompleteness at low redshifts is minor, for the lowest redshift redshift slices this is not a worry, but for
the higher redshift slices the samples are relatively incomplete and might be biased. This will have to be investigated
further using the Phase 1 dataset, where we will also use other methods to estimate the spatial clustering length.

Finally, we found an interesting structure in the redshift cone, at $z=0.164$, which is a likely filamentary
structure roughly aligned with the plane of the sky. It is also seen to have several galaxy groups in the optical
waveband which are absent in the sample we selected for the clustering analysis.
The sources in these groups will provide excellent targets for follow-up
studies with instruments such as ALMA that have much higher spatial resolution than SPIRE.

\section*{Acknowledgements} 

The \herschel-ATLAS is a project with \herschel, which is an ESA space observatory with science instruments
provided by European-led Principal Investigator consortia and with important participation from NASA.
The H-ATLAS website is {\tt http://www.h-atlas.org/}.
GAMA is a joint European-Australasian project based around a spectroscopic campaign using the Anglo-Australian
Telescope. The GAMA input catalogue is based on data taken from the Sloan Digital Sky Survey and the UKIRT
Infrared Deep Sky Survey. Complementary imaging of the GAMA regions is being obtained by a number of independent
survey programs including GALEX MIS, VST KIDS, VISTA VIKING, WISE, \herschel-ATLAS, GMRT and ASKAP, providing
UV to radio coverage. GAMA is funded by the STFC (UK), the ARC (Australia), the AAO, and the participating
institutions. The GAMA website is: {\tt http://www.gama-survey.org/}. 
This research was supported in part by the Austrian Science Foundation FWF 
under grants P18493 and I164, and received financial contribution
from the agreement ASI-INAF I/009/10/0.
Thanks to Duncan Farrah for making his Limber equation inversion code available.


\bibliographystyle{mn2e}

\bsp

\end{document}